\documentclass{kais}
\usepackage{wrapfig,psfig,epsf}
\usepackage[dcucite]{harvard}

\usepackage{graphicx}
\usepackage{color}
\usepackage{longtable}

\received{xxx}
\revised{xxx}
\accepted{xxx}

\pubyear{2000}
\pagerange{\pageref{firstpage}--\pageref{lastpage}}
\volume{xxx}

\begin{document}
\label{firstpage}

\title{Categorization of interestingness measures for knowledge extraction}

\author[Guillaume et al]{Sylvie Guillaume$^1$ $^3$, Dhouha Grissa$^2$ $^3$ $^4$, Engelbert Mephu Nguifo$^2$ $^3$\\
~~Clermont Universit\'e, Universit\'e d'Auvergne $^1$, Universit\'e Blaise Pascal $^2$,\\ 
~~LIMOS, F-63000 CLERMONT-FERRAND\\
%$^2$ Clermont Universit\'e, Universit\'e Blaise Pascal, LIMOS, F-63000 CLERMONT-FERRAND\\
$^3$ CNRS, UMR 6158, LIMOS, F-63173 AUBIERE\\
$^4$ LIPAH, D\'epartement d'Informatique, Facult\'e des Sciences de Tunis,\\
~~Campus Universitaire, 1060 TUNIS, TUNISIE \\
~~\texttt{\{dgrissa,guillaum,mephu\}@isima.fr}\\ 
}

 %$^1$Department of Computer Science, Dartmouth
%College, Hanover NH, USA; 
%$^2$School of Computing\\ and Information
%Technology, Griffith University, Nathan, Queensland, Australia}

\maketitle

\begin{abstract}

Finding interesting association rules is an important and active research field in data
mining. The algorithms of the {\em Apriori\/} family are based on two rule extraction measures,
support and confidence. Although these two measures have the virtue of being
algorithmically fast, they generate a prohibitive number of rules most of which are redundant
and irrelevant. It is therefore necessary to use further measures which filter uninteresting
rules. Many synthesis studies were then realized on the interestingness measures according to several points of view. Different reported studies have been carried out to identify "\textit{good}" properties of rule
extraction measures and these properties have been assessed on $61$ measures. The purpose of this paper is twofold. First to extend the number of the measures and properties to be studied, in addition to the formalization of the properties proposed in the literature. Second, in the light of this formal study, to categorize the studied measures. This paper leads then to identify categories of measures in order to help the users to efficiently select an appropriate measure by choosing one or more measure(s) during the knowledge extraction process. The properties evaluation on the $61$ measures has enabled us to identify $7$ classes of measures, classes that we obtained using two different clustering techniques.

%To help carry out the task, many measures which evaluate the interestingness of rules have been developed. The data mining experts should choose an appropriate interestingness measure in order to filter the huge amount of rules. Nevertheless, as this study shows, this choice is not easy
%a two-step solution to the problem of the   . First,

\end{abstract}


\section{Introduction}
\label{sec:intro}

%{\sf minconf\/} bold

%{\em \/} italic

Association rules mining algorithms \cite{Agrawal94}, based on support and confidence measures, tend to generate a large number of rules. These two measures are not sufficient to extract only the really interesting rules and this statement was highlighted in many studies such as \cite{Sese02}, \cite{Carvalho05}. An additional step of analyzing extracted rules is therefore essential and different solutions have been proposed. A first solution consists of restoring easily and with a synthetic way, the extracted information through visual representation techniques \cite{Hof01}, \cite{BlanchardGB03}. A second way is to reduce the number of rules. Some authors \cite{Zaki00}, \cite{Zaman04}, \cite{YahiaGN09} eliminate redundant rules, others evaluate and order the rules due to some interestingness measures \cite{LencaMVL08}. In this paper, we focus on the latter path: the use of interestingness measures to eliminate uninteresting rules. Many synthesis studies compared the different objective measures reported in the literature according to several points of view: underlying properties for a "\textit{good}" interestingness measure \cite{Tan02}, \cite{lallich04}, \cite{Vaillant06}, \cite{GengH07:crl_fwidm}, \cite{Feno07}, \cite{HeraviZ10}. These synthetic articles highlighted some of the interestingness measures reported in the literature with some of the proposed properties.

The purpose of this paper is twofold: first to extend the number of the measures and properties to be studied, in addition to the formalization of the different properties proposed in the literature; and second, in the light of this formal study which is performed by the evaluation of interestingness measures according to "\textit{good}" properties, to categorize the studied measures and to interpret the detected classes. We then wish to detect groups of measures with similar properties, allowing the user from one hand, to restrict the number of measures to choose from, and secondly, to direct his choice based on the properties he wishes that measures check.

% assist the user in selecting one or more additional measure(s) to eliminate irrelevant \footnote{The relevance or the rule interest is measured with respect to the studied problem, and some relevant rules may not to be valid because of the used measure.}  rules extracted by the couple (\textit{Support, Confidence}). For this, we wish 

%This work is based on the synthetic works that have been made on interestingness measures and their properties and especially on the work of Guillaume et al. \cite{Guillaume10}) as being the most recent article in this area, it is the most comprehensive since it is a synthesis of \cite{Tan02}, \cite{lallich04}, \cite{HuynhGB05:cimpc}, \cite{Vaillant06}, \cite{GengH07:crl_fwidm}, \cite{Feno07} works. This synthesis study of Guillaume et al. \cite{Guillaume10} has listed some sixty interestingness measures and twenty properties. This work culminated with the evaluation of $19$ properties on $61$ measures.

Therefore, we want to check classes of measures with similar behavior compared to all the properties we have identified but in any case to explain the properties and measures identified in the literature, explanations can be found in review articles \cite{Tan02}, \cite{lallich04}, \cite{Vaillant06}, \cite{GengH07:crl_fwidm}, \cite{Feno07}. The search for these classes of measures was performed using well known techniques as one of the methods of agglomerative hierarchical clustering using Ward criterion \cite{Ward:hgobf} and a version of a non-hierarchical clustering method of \textit{k-means} \cite{Mac67}. A consensus is then derived from the results obtained with both techniques. Before starting the search for classes, it became essential to check that this matrix of \textit{measures} $\times$ \textit{properties} could not be simplified by looking for groups of measures with completely similar behavior in relation to all the \textit{properties} and also, if there was no redundant properties.

The article is thus organized as follows. \textit{Section} \ref{sec:recprop} presents and formalizes the different properties. \textit{Section} \ref{sec:evalprop} outlines the matrix of $ measures \times properties$ on which we look for classes and studying if it can not be simplified. \textit{Section} \ref{sec:clacah} restitutes the results of the classification obtained by the first technique: a method of agglomerative hierarchical clustering using Ward criterion. \textit{Section} \ref{sec:clakmfi} gives the results generated by the second technique: a version of the non-hierarchical clustering method of \textit{k-means} and discusses the consistency of the results obtained by both techniques. The section ends with a consensus classification. Finally, \textit{Section} \ref{sec:clreva} tries to find a semantic to some of the extracted classes and valid the retained classification to those released by \cite{Vaillant06}, \cite{LeBras2011}, \cite{HuynhGB07}, \cite{Lesot:2010:OED}, \cite{ZighedAB11}. The article ends with a conclusion and perspectives.

%%%%%%%%%%%%%%%%%%%%%%%%%%%%%%%%%%%%%%%%%%

\section{Association rules}\label{sec:ar}

As defined in \cite{Agrawal93}, given \textit{I = \{$i_{1},...,i_{n}$\}} be a set of \textit{k} items and \textit{B = \{$b_{1},...,b_{n}$\}} a basket database representing a collection of \textit{n} subset of items \textit{I}, an \emph{association rule} \cite{Agrawal93} in the database \textit{B} is a formula

\begin{center}
%\begin{equation}
$  X\Rightarrow Y$
%\end{equation}
\end{center}

where $X$ and $Y$ are sets of items from $I$, i.e. $X,Y\subseteq I$ with $X \cap Y = \emptyset$. $X$ represents the antecedent or premise of this rule and $Y$ the consequent or conclusion.

A natural interestingness measure of association rules 
is based on the notions of support and confidence.
The \emph{Support} (\textit{when X and Y occur together in at least $s\%$ of the n baskets}) and \emph{Confidence} (\textit{when from all the baskets containing X, at least $c\%$ also contain Y}) of an association rule $X\Rightarrow Y$ are defined by

\[
     \textrm{Supp}(X\Rightarrow Y) = Supp(X \cup Y)
   \quad\mbox{and}\quad
    \textrm{Conf}(X\Rightarrow Y) = \frac{Supp(X \cup Y)}{Supp(X)},
\]

An association rule is considered interesting if its confidence and support exceed some user-specified thresholds. 

However, the support-confidence approach reveals some weaknesses. Often, this approach as well 
as algorithms based on it lead to the extraction of an exponential number of rules. 
Therefore, it is impossible to validate it by an expert. In addition, the disadvantage of the support 
is that sometimes many rules that are potentially interesting, have a lower support value and therefore 
can be eliminated by the pruning threshold \textit{minsupp}. To address this problem, many other measures 
of interestingness have been proposed in the literature~\cite{GengH07:crl_fwidm}, mainly because they are 
effective for mining potentially interesting rules and 
capture some aspects of user interest. The most important of those measures
are subject to our analysis and are surveyed in \textit{Annexe} of \textit{section} \ref{sec:annexe1}. However, the concept
of association rule itself as well as various measures of interestingness are particular
cases of what is investigated in depth in \cite{HaHa:MHF}, a book that develops
logico-statistical foundations of the GUHA method \cite{HaHoRa:Gmmdm}.

%An example of such an association rule is the statement that $90\%$ of transactions that purchase bread and butter also purchase milk. The antecedent of this rule consists of bread and butter and the consequent consists of milk alone. The number $90\%$ is the confidence factor of the rule.

\section{Recall and formalization of the properties}
\label{sec:recprop}

The following section presents the different properties of measures reported in the literature. We then recall these properties afterward we formalize them for a better understanding.

%\subsection{Properties of measures}
%\label{subsec:prop}

This section describes the properties currently used in the literature to characterize measures. Those properties are then summarized in \textit{table} \ref{prop_meas}.\\

\begin{table}
\centering
\caption{Properties of measures {\em m\/}}
\begin{tabular}{|p{0,5cm}|p{6,9cm}|} \hline

\textbf{$N^{\circ}$}&\textbf{Properties} \\ \hline

\textcolor{blue}{$P_{1}$} & The measure {\em m\/} is Asymmetric (\textcolor{blue}{$P_{1}(m) = 1$}) or symmetric (\textcolor{blue}{$P_{1}(m) = 0$}). \\\hline

\textcolor{blue}{$P_{2}$} & {\em m\/} does not equalize the antinomic rules (\textcolor{blue}{$P_{2}(m) = 1$}) or equalizes them (\textcolor{blue}{$P_{2}(m) = 0$}). \\\hline

\textcolor{blue}{$P_{3}$} & {\em m\/} assesses in the same way the rules $X \rightarrow Y$ and $\bar{Y} \rightarrow \bar{X}$ in the logical implication case (\textcolor{blue}{$P_{3}(m) = 1$}) or not (\textcolor{blue}{$P_{3}(m) = 0$}). \\\hline

\textcolor{blue}{$P_{4}$} & {\em m\/} increases according to the number of examples (\textcolor{blue}{$P_{4}(m) = 1$}) or decreases (\textcolor{blue}{$P_{4}(m) = 0$}).\\\hline

\textcolor{blue}{$P_{5}$} & {\em m\/} increases according to the size of the training set (\textcolor{blue}{$P_{5}(m) = 1$}) or not (\textcolor{blue}{$P_{5}(m) = 0$}). \\\hline

\textcolor{blue}{$P_{6}$} & {\em m\/} decreases according to the consequent size (\textcolor{blue}{$P_{6}(m) = 1$}) or increases (\textcolor{blue}{$P_{6}(m) = 0$}).\\\hline

\textcolor{blue}{$P_{7}$} & {\em m\/} has a fixed value in the independence case (\textcolor{blue}{$P_{7}(m) = 1$}) or not (\textcolor{blue}{$P_{7}(m) = 0$}).\\\hline

\textcolor{blue}{$P_{8}$} & {\em m\/} has a fixed value in the logical implication case (\textcolor{blue}{$P_{8}(m) = 1$}) or not (\textcolor{blue}{$P_{8}(m) = 0$}). \\\hline

\textcolor{blue}{$P_{9}$} & {\em m\/} has a fixed value in the equilibrium case (\textcolor{blue}{$P_{9}(m) = 1$}) or not (\textcolor{blue}{$P_{9}(m) = 0$}).\\\hline

\textcolor{blue}{$P_{10}$} & Identified values in the attraction case between $X$ and $Y$ (\textcolor{blue}{$P_{10}(m) = 1$}) or not (\textcolor{blue}{$P_{10}(m) = 0$}). \\\hline

\textcolor{blue}{$P_{11}$} & Identified values in the repulsion case between $X$ and $Y$ (\textcolor{blue}{$P_{11}(m) = 1$}) or not (\textcolor{blue}{$P_{11}(m) = 0$}). \\\hline

\textcolor{blue}{$P_{12}$} & {\em m\/} is tolerant to the first counter-examples (\textcolor{blue}{$P_{12}(m) = 2$}) or not tolerant (\textcolor{blue}{$P_{12}(m) = 0$}) or indifferent (\textcolor{blue}{$P_{12}(m) = 1$}).\\\hline

\textcolor{blue}{$P_{13}$} & {\em m\/} invariant in case of expansion of certain quantities (\textcolor{blue}{$P_{13}(m) = 1$}) or not (\textcolor{blue}{$P_{13}(m) = 0$}). \\\hline

\textcolor{blue}{$P_{14}$} & {\em m\/} opposes the rules $X \rightarrow Y$ and $\bar{X} \rightarrow Y$ (\textcolor{blue}{$P_{14}(m) = 1$}) or not (\textcolor{blue}{$P_{14}(m) = 0$}). \\\hline

\textcolor{blue}{$P_{15}$} & {\em m\/} oppposes the antinomic rules $X \rightarrow Y$ and $X \rightarrow \bar{Y}$ (\textcolor{blue}{$P_{15}(m) = 1$}) or not (\textcolor{blue}{$P_{15}(m) = 0$}).\\\hline

\textcolor{blue}{$P_{16}$} & {\em m\/} equalizes the rules $X \rightarrow Y$ and $\bar{X} \rightarrow \bar{Y}$ (\textcolor{blue}{$P_{16}(m) = 1$}) or not (\textcolor{blue}{$P_{16}(m) = 0$}). \\\hline

\textcolor{blue}{$P_{17}$} & {\em m\/} is based on a probabilistic model (\textcolor{blue}{$P_{17}(m) = 1$}) or not (\textcolor{blue}{$P_{17}(m) = 0$}). \\\hline

\textcolor{blue}{$P_{18}$} & {\em m\/} is statistic (\textcolor{blue}{$P_{18}(m) = 1$}) or descriptive (\textcolor{blue}{$P_{18}(m) = 0$}).\\\hline

\textcolor{blue}{$P_{19}$} & {\em m\/} is discriminant (\textcolor{blue}{$P_{19}(m) = 1$}) or not (\textcolor{blue}{$P_{19}(m) = 0$}). \\

\hline\end{tabular}
\label{prop_meas}
\end{table}

We give some details about the terminology given in \textit{table} \ref{prop_meas}:

\begin{itemize}

\item \textbf{Example}: individual who checks both the premise {\em X\/} and the conclusion {\em Y\/} of the rule,
\item \textbf{Independence}: case where the realisation of {\em X\/} does not increase the chances of occurrence of {\em Y\/},
\item \textbf{Logical implication}: if the conditional probability {\em P(Y/X)\/} is equal to $1$,
\item \textbf{Equilibrium or indetermination}: case where {\em Y\/} is achieved when there is much chance that {\em X\/} or {\em not X\/} be realized,
\item \textbf{Attraction}: when the realization of {\em X\/} increases the chances of occurrence of {\em Y\/},
\item \textbf{Repulsion}: when the realization of {\em X\/} decreases the chances of occurrence of {\em Y\/}.

\end{itemize}

%After this description of measures properties, the next section formalizes them.
%
%\subsection{Properties formalization}
%\label{subsec:formprop}

We formalize the different properties encountered in the literature and exposed in \textit{table} \ref{prop_meas}. The title of the $21$ properties listed is, preferably, the desired property for a measure $m$.\\

{\bf \textcolor{blue}{Property 1} : Asymmetric measure.}

\begin{center}
%\begin{tabular}{|l|}
\fcolorbox[RGB]{230,230,250}{245,245,245}{\parbox{\linewidth}{

{\footnotesize  $P_{1}(m)~=~ 0~~ if~~ m~ is~ symmetric ~~~~~~~~i.e~~ if ~\forall~ X~ \rightarrow ~Y~~~~~~m(X~ \rightarrow~ Y)~ =~ m(Y~ \rightarrow~ X)$} \\
{\footnotesize  $P_{1}(m)~=~ 1~~ if~~ m~ is~not~ symmetric ~~i.e~~ if ~~\exists~ X~ \rightarrow ~Y~~~/~m(X~ \rightarrow~ Y)~ \neq~ m(Y~ \rightarrow~ X)$} \\
}}
%\end{tabular}
\end{center}

\vspace{0.4cm}
{\bf \textcolor{blue}{Property 2} : Asymmetric measure in the sense of the conclusion negation or measure does not equalize the antinomic rules}

\begin{center}
%\begin{tabular}{|l|}
\fcolorbox[RGB]{230,230,250}{255,239,213}{\parbox{\linewidth}{

{\footnotesize  $P_{2}(m)~=~ 0~~ if~~ m~ is~cn-symmetric ~~~~~~~i.e~~ if ~~ \forall~ X~ \rightarrow ~Y~~~~~m(X~ \rightarrow~ Y)~ =~ m(X~ \rightarrow~ \bar{Y})$} \\
{\footnotesize  $P_{2}(m)~=~ 1~~ if~~ m~ is~ not~cn-symmetric ~~i.e~~ if~~ \exists~ X~ \rightarrow ~Y~~~/~m(X~ \rightarrow~ Y)~ \neq~ m(X~ \rightarrow~\bar{Y})$} \\

}}
%\end{tabular}
\end{center}

\vspace{0.4cm}
{\bf \textcolor{blue}{Property 3} : Measure assessing in the same way $X \rightarrow Y$ and $\bar{Y} \rightarrow \bar{X}$ in the logical implication case.}

\begin{center}
%\begin{tabular}{|l|}
\fcolorbox[RGB]{230,230,250}{245,245,245}{\parbox{\linewidth}{
{\footnotesize $P_{3}(m)~=~ 0~~ if~~ \exists~ X~ \rightarrow ~Y~/~~P(Y/X)=1 ~~and~ ~m(X~ \rightarrow~ Y)~ \neq~ m(\bar{Y}~ \rightarrow~ \bar{X})$} \\
{\footnotesize  $P_{3}(m)~=~ 1~~ if~~ \forall~ X~ \rightarrow~Y~~~~P(Y/X)=1~~ \Rightarrow~~m(X~ \rightarrow~ Y)~ =~ m(\bar{Y}~ \rightarrow~\bar{X})$ }\\ 

}}
%\end{tabular}
\end{center}

\vspace{0.4cm}
{\bf \textcolor{blue}{Property 4} : Measure increasing according to the number of examples or decreasing with the number of counter-examples the number of records satisfying \textit{X} but not \textit{Y}.}.

\begin{center}
%\begin{tabular}{|l|}
\fcolorbox[RGB]{230,230,250}{255,239,213}{\parbox{\linewidth}{

{\footnotesize  $P_{4}(m)~=~ 0~~ if~~m~didn't~increase~with~n_{XY}~i.e.~~if~~\exists~ X_{1}~ \rightarrow ~Y_{1},~~ \exists~ X_{2}~ \rightarrow ~Y_{2}/ $ }\\ 
{\footnotesize $n_{X_{1}}=n_{X_{2}}~~and~~n_{Y_{1}}=n_{Y_{2}} ~~and~~(n_{X_{1}}n_{Y_{1}}<n_{X_{2}}n_{Y_{2}}~or~n_{X_{1}}n_{\bar{Y}_{1}}>n_{X_{2}}n_{\bar{Y}_{2}}) $}\\
{\footnotesize  $~~~~~~~~~~~~~~~~~~~~~~~~~~~~~and~~m(X_{1}~ \rightarrow ~Y_{1}) \geq m(X_{2}~ \rightarrow ~Y_{2}), $ }\\

{\footnotesize $P_{4}(m)~=~ 1~~ if~~m~is~increasing~with~n_{XY}~i.e.~~if~~\forall~ X_{1}~ \rightarrow ~Y_{1},~\forall~ X_{2}~ \rightarrow ~Y_{2}$ }\\
{\footnotesize $~[n_{X_{1}}=n_{X_{2}}~~and~~n_{Y_{1}}=n_{Y_{2}}~~and~~(n_{X_{1}Y_{1}}<n_{X_{2}Y_{2}}~~or~n_{X_{1} \bar{Y}_{1}}>n_{X_{2}\bar{Y}_{2}})] $}\\
{\footnotesize $~~~~~~~~~~~~~~~~~~~~~~~~~~~~~~ \Rightarrow~m(X_{1}~ \rightarrow ~Y_{1}) \leq m(X_{2}~ \rightarrow ~Y_{2})~~and $}\\
{\footnotesize $ [\exists~ X_{1}~ \rightarrow ~Y_{1},~~ \exists~ X_{2}~ \rightarrow ~Y_{2}]~/~n_{X_{1}}=n_{X_{2}}~~and~~n_{Y_{1}}=n_{Y_{2}}~~and~~(n_{X_{1}Y_{1}}~ <~ n_{X_{2}Y_{2}}$}\\
{\footnotesize $ or~n_{X_{1} \bar{Y}_{1}} ~>~ n_{X_{2}\bar{Y}_{2}})~~ and~~ m(X_{1}~ \rightarrow ~Y_{1}) < m(X_{2}~ \rightarrow ~Y_{2})] $} \\

}}
%\end{tabular}
\end{center}

With $n_{XY} = \vert X \cap Y \vert$ the number of records satisfying both \textit{X} and \textit{Y} and $n_{X\bar{Y}} = \vert X \cap \bar{Y} \vert$.

\vspace{0.4cm}
{\bf \textcolor{blue}{Property 5} : Measure increasing according to the size of the training set $n$}

\begin{center}
%\begin{tabular}{|l|}
\fcolorbox[RGB]{230,230,250}{245,245,245}{\parbox{\linewidth}{

{\footnotesize $P_{5}(m)~=~ 0~~ (m~didn't~increase~with~n)~~if~~\exists~(\Omega_{1},~ \Omega_{2}), $ }\\
{\footnotesize $~\exists~ X_{1}~ \rightarrow ~Y_{1}~(\Omega_{1}),~\exists~ X_{2}~ \rightarrow ~Y_{2}~(\Omega_{2})~/~n_{X_{1}}=n_{X_{2}}~~and~~n_{Y_{1}}=n_{Y_{2}} $ }\\~ {\footnotesize  $and~~n_{X_{1}Y_{1}}=n_{X_{2}Y_{2}}~~and~~n_{1}<n_{2} ~~~~and~~~m(X_{1}~ \rightarrow ~Y_{1}) > m(X_{2}~ \rightarrow ~Y_{2})$} \\
%\vspace{0.1cm}
{\footnotesize $P_{5}(m)~=~ 1~~ (m~increases~with~n)~~if~~\forall~\Omega_{1},~\forall~\Omega_{2},$}\\
{\footnotesize $~~\forall~ X_{1}~ \rightarrow ~Y_{1}~(\Omega_{1}),~\forall~ X_{2}~ \rightarrow ~Y_{2}~(\Omega_{2})~~(n_{X_{1}}=n_{X_{2}}~~and~~n_{Y_{1}}=n_{Y_{2}} $ }\\
 {\footnotesize $and~~n_{X_{1}Y_{1}}=n_{X_{2}Y_{2}}~~and~~n_{1}<n_{2}) ~~~~\Rightarrow~~~m(X_{1}~ \rightarrow ~Y_{1}) \leq m(X_{2}~ \rightarrow ~Y_{2})$ }\\
{\footnotesize $and~~\exists~\Omega_{1},~\exists~\Omega_{2},~\exists~ X_{1}~ \rightarrow ~Y_{1}~(\Omega_{1}),~\exists~ X_{2}~ \rightarrow ~Y_{2}~(\Omega_{2})~/$}\\
{\footnotesize $~(n_{X_{1}}=n_{X_{2}}~~and~~n_{Y_{1}}=n_{Y_{2}}~~and~~n_{X_{1}Y_{1}}=n_{X_{2}Y_{2}}~~and~~n_{1}<n_{2}~) $}\\
{\footnotesize $~~~~~~~~~~~~~~~~~~~~~~~~~~~~~~~~~~~~~~~~~~~~~~~~~~~and~~~m(X_{1}~ \rightarrow ~Y_{1}) < m(X_{2}~ \rightarrow ~Y_{2})$ }\\

}}
%\end{tabular}
\end{center}

\vspace{0.4cm}
{\bf \textcolor{blue}{Property 6} : Measure decreasing according to the the size of the consequent \footnote{$n_{Y} = \vert Y \vert$ the number of records satisfying \textit{Y}.} or the size of the premise \footnote{$n_{X} = \vert X \vert$ the number of records satisfying \textit{X}.}.}

\begin{center}
%\begin{tabular}{|l|}
\fcolorbox[RGB]{230,230,250}{255,239,213}{\parbox{\linewidth}{

{\footnotesize $P_{6}(m)~=~ 0~~ if~~m~didn't~decrease~with~n_{Y}~~i.e.~~if$ }\\
{\footnotesize $~~\exists~ X_{1}~ \rightarrow ~Y_{1},~\exists~ X_{2}~ \rightarrow ~Y_{2}~/~n_{X_{1}}=n_{X_{2}}~ and~n_{X_{1}Y_{1}}=n_{X_{2}Y_{2}}~and~n_{Y_{1}}<n_{Y_{2}} $}\\
{\footnotesize $~~~~~~~~~~~~~~~~~~~~~~~~~and~~m(X_{1}~ \rightarrow ~Y_{1}) < m(X_{2}~ \rightarrow ~Y_{2}), $} \\

{\footnotesize $P_{6}(m)~=~ 1~~ if~~m~is~decreasing~with~n_{Y}~~i.e.~~if $}\\
{\footnotesize $~~\forall~ X_{1}~ \rightarrow ~Y_{1},~\forall~ X_{2}~ \rightarrow ~Y_{2}~(n_{X_{1}}=n_{X_{2}}~ and~n_{X_{1}Y_{1}}=n_{X_{2}Y_{2}}~and~n_{Y_{1}}<n_{Y_{2}}) $}\\
{\footnotesize $ ~~~~~~~~~~~~~~~~~~~~~~~~~\Rightarrow~~m(X_{1}~ \rightarrow ~Y_{1}) \geq m(X_{2}~ \rightarrow ~Y_{2})~and, $ }\\
{\footnotesize $~~\exists~ X_{1}~ \rightarrow ~Y_{1},~\exists~ X_{2}~ \rightarrow ~Y_{2}/(n_{X_{1}}=n_{X_{2}}~ and~n_{X_{1}Y_{1}}=n_{X_{2}Y_{2}}~and~n_{Y_{1}}<n_{Y_{2}}) $}\\
{\footnotesize $ ~~~~~~~~~~~~~~~~~~~~~~~~~and~~m(X_{1}~ \rightarrow ~Y_{1}) > m(X_{2}~ \rightarrow ~Y_{2})$ }\\

}}
%\end{tabular}
\end{center}

If we consider the premise size, the property $P_{6}(m)=1$ is also written:

\vspace{0.4cm}

\begin{center}
%\begin{tabular}{|l|}
\fcolorbox[RGB]{230,230,250}{255,239,213}{\parbox{\linewidth}{

{\footnotesize $P_{6}(m)~=~ 1~~ if~m~is~decreasing~with~n_{x}~~i.e.~when $}\\
{\footnotesize $~~\forall~ X_{1}~ \rightarrow ~Y_{1},~\forall~ X_{2}~ \rightarrow ~Y_{2}~(n_{Y_{1}}=n_{Y_{2}}~ and~n_{X_{1}Y_{1}}=n_{X_{2}Y_{2}}~and~n_{X_{1}}<n_{X_{2}}) $}\\
{\footnotesize $ ~~~~~~~~~~~~~~~~~~~~~~~~~\Rightarrow~~m(X_{1}~ \rightarrow ~Y_{1}) > m(X_{2}~ \rightarrow ~Y_{2})$ }\\
}}
%\end{tabular}
\end{center}

\vspace{0.4cm}
{\bf \textcolor{blue}{Property 7} : Fixed value $a$ in the independence case.}

\begin{center}
%\begin{tabular}{|l|}
\fcolorbox[RGB]{230,230,250}{245,245,245}{\parbox{\linewidth}{

{\footnotesize $P_{7}(m)~=~ 0~~~~~~~~~~~~~~~~~~~~~ if~~\forall~a \in \mathbf{R}~~~ \exists~~ X~ \rightarrow ~Y~ /~P(Y/X)=P(Y)$} \\
{\footnotesize $~~~~~~~~~~~~~~~~~~~~~~~~and~~m(X~ \rightarrow~ Y)~ \neq~a$} \\
{\footnotesize $P_{7}(m)~=~ 1~~(fixed~value)~ if~~\exists~a \in \mathbf{R}~/~\forall~ X~ \rightarrow~Y~~~P(Y/X)=P(Y)$} \\ 
{\footnotesize $~~~~~~~~~~~~~~~~~~~~~~~~~\Rightarrow~m(X~ \rightarrow~ Y)~ =~a$} \\

}}
%\end{tabular}
\end{center}

\vspace{0.4cm}
{\bf \textcolor{blue}{Property 8} : Fixed value $b$ in the logical implication case.}

\begin{center}
%\begin{tabular}{|l|}
\fcolorbox[RGB]{230,230,250}{255,239,213}{\parbox{\linewidth}{

{\footnotesize $P_{8}(m)~=~ 0~~~~~~~~~~~~~~~~~~~~~  if~~\forall~b \in \mathbf{R}~~~ \exists~~ X~ \rightarrow ~Y~ /~P(Y/X)=1$ }\\
{\footnotesize $~~~~~~~~~~~~~~~~~~~~~~~~and~~m(X~ \rightarrow~ Y)~ \neq~b$} \\
{\footnotesize $P_{8}(m)~=~ 1~~(fixed~value)~if~~\exists~b \in \mathbf{R}~/~\forall~ X~ \rightarrow ~Y~~~ ~P(Y/X)=1$} \\ 
{\footnotesize $~~~~~~~~~~~~~~~~~~~~~~~~~\Rightarrow~m(X~ \rightarrow~ Y)~ =~b$} \\
}}
%\end{tabular}
\end{center}

\vspace{0.4cm}
{\bf \textcolor{blue}{Property 9} : Fixed value $c$ in the equilibrium case.}

\begin{center}
%\begin{tabular}{|l|}
\fcolorbox[RGB]{230,230,250}{245,245,245}{\parbox{\linewidth}{

{\footnotesize $P_{9}(m)~=~ 0~~~~~~~~~~~~~~~~~~~~~  if~~\forall~c \in \mathbf{R}~~~ \exists~~ X~ \rightarrow ~Y~ /~P(Y/X)=P(X)/2$ }\\
{\footnotesize $~~~~~~~~~~~~~~~~~~~~~~~~and~~m(X~ \rightarrow~ Y)~ \neq~c$} \\
{\footnotesize $P_{9}(m)~=~ 1~~(fixed~value)~ if~~\exists~c \in \mathbf{R}~/~\forall~ X~ \rightarrow ~Y~~~ ~P(Y/X)=P(X)/2 $} \\ 
{\footnotesize $~~~~~~~~~~~~~~~~~~~~~~~~~\Rightarrow~m(X~ \rightarrow~ Y)~ =~c$} \\

}}
%\end{tabular}
\end{center}

\vspace{0.4cm}
{\bf \textcolor{blue}{Property 10} : Identified values in the attraction case between $X$ and $Y$.}

\begin{center}
%\begin{tabular}{|l|}
\fcolorbox[RGB]{230,230,250}{255,239,213}{\parbox{\linewidth}{
{\footnotesize $P_{10}(m)~=~ 0~~~~~~~~~~~~~~~~~~~~~~~~~~~~~if~~\forall~a \in \mathbf{R}~~~ \exists~ X~ \rightarrow ~Y~ /~P(Y/X)>P(Y)$ }\\
{\footnotesize $~~~~~~~~~~~~~~~~~~~~~~~~and~~m(X~ \rightarrow~ Y)~ \leq~a$} \\
{\footnotesize $P_{10}(m)~=~ 1~~(identified~values)~ if~~\exists~a \in \mathbf{R}~/~ \forall~ X~ \rightarrow ~Y~~ ~P(Y/X)>P(Y)$} \\ 
{\footnotesize $~~~~~~~~~~~~~~~~~~~~~~~~~\Rightarrow~m(X~ \rightarrow~ Y)~ >~a$} \\
}}
%\end{tabular}
\end{center}

\vspace{0.4cm}
{\bf \textcolor{blue}{Property 11} : Identified values in the repulsion case between $X$ and $Y$.}

\begin{center}
%\begin{tabular}{|l|}
\fcolorbox[RGB]{230,230,250}{245,245,245}{\parbox{\linewidth}{
{\footnotesize $P_{11}(m)~=~ 0~~~~~~~~~~~~~~~~~~~~~~~~~~~~~ if~~\forall~a \in \mathbf{R}~~~ \exists~ X~ \rightarrow ~Y~ /~P(Y/X) < P(Y)$ }\\
{\footnotesize $~~~~~~~~~~~~~~~~~~~~~~~~and~~m(X~ \rightarrow~ Y)~ \geq~a$} \\
{\footnotesize $P_{11}(m)~=~ 1~~ (identified~values)~ if~~\exists~a \in \mathbf{R}~/~ \forall~ X~ \rightarrow ~Y~~ ~P(Y/X) < P(Y)$} \\ 
{\footnotesize $~~~~~~~~~~~~~~~~~~~~~~~~~\Rightarrow~m(X~ \rightarrow~ Y) < a$} \\

}}
%\end{tabular}
\end{center}

\vspace{0.4cm}
{\bf \textcolor{blue}{Property 12} : Tolerance to the first counter-examples.}

\begin{center}
%\begin{tabular}{|l|}
\fcolorbox[RGB]{230,230,250}{255,239,213}{\parbox{\linewidth}{
{\footnotesize
$P_{12}(m)~=~ 0~~ if~reject~so~convex,~\exists~min_{conf} \in [0,1] / \forall~ X_{1}~ \rightarrow ~Y_{1},~\forall~ X_{2}~ \rightarrow ~Y_{2}$}\\
{\footnotesize $~\forall~\lambda \in [0,1]~~n_{X_{1}Y_{1}} \geq min_{conf}~ n(X_{1})~~and~~ n_{X_{2}Y_{2}} \geq min_{conf}~ n(X_{2}) $}\\
%\vspace{0.1cm} 
{\footnotesize $~~~\Rightarrow~~f_{m, n_{XY}}(\lambda n_{X_{1}Y_{1}} + (1-\lambda)n_{X_{2}Y_{2}}) \leq \lambda f_{m, n_{XY}} (n_{X_{1}Y_{1}}) + (1-\lambda) f_{m, n_{XY}}(n_{X_{2}Y_{2}})$}\\

{\footnotesize $P_{12}(m)~=~ 1~~ if~indifference~then~linear~~i.e.~~P_{14}(m)\neq 0 ~~and~~ P_{14}(m)\neq 2 $}\\

{\footnotesize $P_{12}(m)~=~ 2~~ if~tolerance~then~concave~\exists~min_{conf} \in [0,1] /\forall~X_{1}~ \rightarrow ~Y_{1},\forall~ X_{2}~ \rightarrow ~Y_{2},$ }\\
{\footnotesize $\forall~\lambda \in  [0,1]~~n_{X_{1}Y_{1}}  \geq  min_{conf}~n(X_{1})~~and~~ n_{X_{2}Y_{2}} \geq min_{conf}~n(X_{2}) $}\\
%\vspace{0.1cm} 
{\footnotesize $~~~\Rightarrow~~f_{m, n_{XY}}(\lambda n_{X_{1}Y_{1}} + (1-\lambda)n_{X_{2}Y_{2}}) \geq \lambda f_{m, n_{XY}} (n_{X_{1}Y_{1}}) + (1-\lambda) f_{m, n_{XY}}(n_{X_{2}Y_{2}})$}\\

}}
%\end{tabular}
\end{center}

\vspace{0.25cm}

The notation $f_{m, n_{XY}}$ corresponds to the evolution according to the measure $m$ with $n_{XY}$ when $n_{X}$, $n_{Y}$ and $n$ remain constant.

\vspace{0.4cm}
{\bf \textcolor{blue}{Property 13} : Invariance in case of expansion of certain quantities ($n_{XY}$, $n_{\bar{X}Y}$ and $n_{X\bar{Y}}$).}

\begin{center}
%\begin{tabular}{|l|}
\fcolorbox[RGB]{230,230,250}{245,245,245}{\parbox{\linewidth}{
{\footnotesize
$P_{13}(m)~=~ 0~~(variance)~ if~~\exists~(k_{1}, k_{2}) \in N_{*2},~\exists~ X_{1}~ \rightarrow ~Y_{1},~\exists~ X_{2}~ \rightarrow ~Y_{2}~/$}\\
{\footnotesize $[n_{X_{1}Y_{1}} = k_{1}n_{X_{2}Y_{2}} ~ and~n_{X_{1}\bar{Y}_{1}}= k_{1}n_{X_{2}\bar{Y}_{2}}~and ~n_{\bar{X}_{1}Y_{1}} = k_{2}n_{\bar{X}_{2}Y_{2}} ~ and~n_{\bar{X}_{1}\bar{Y}_{1}}= k_{2}n_{\bar{X}_{2}\bar{Y}_{2}}$}\\
%\vspace{0.1cm} 
{\footnotesize $~~~~~~~~~~~~~and~~m(X_{1}~ \rightarrow~ Y_{1}) \neq m(X_{2}~ \rightarrow~ Y_{2})]~or$}\\

{\footnotesize $[n_{X_{1}\bar{Y}_{1}} = k_{1}n_{X_{2}\bar{Y}_{2}} ~ and~n_{\bar{X}_{1}\bar{Y}_{1}}= k_{1}n_{\bar{X}_{2}\bar{Y}_{2}}~and ~n_{X_{1}Y_{1}} = k_{2}n_{X_{2}Y_{2}} ~ and~n_{\bar{X}_{1}Y_{1}}= k_{2}n_{\bar{X}_{2}Y_{2}}$}\\
%\vspace{0.1cm} 
{\footnotesize $~~~~~~~~~~~~~and~~m(X_{1}~ \rightarrow~ Y_{1}) \neq m(X_{2}~ \rightarrow~ Y_{2})]$}\\

{\footnotesize $P_{13}(m)~=~ 1~~(invariance)~ if~~\forall~(k_{1}, k_{2}) \in N_{*2},~\forall~ X_{1}~ \rightarrow ~Y_{1},~\forall~ X_{2}~ \rightarrow ~Y_{2}~/$}\\
{\footnotesize $[(n_{X_{1}Y_{1}} = k_{1}n_{X_{2}Y_{2}} ~ and~n_{X_{1}\bar{Y}_{1}}= k_{1}n_{X_{2}\bar{Y}_{2}}~and ~n_{\bar{X}_{1}Y_{1}} = k_{2}n_{\bar{X}_{2}Y_{2}} ~ and~n_{\bar{X}_{1}\bar{Y}_{1}}= k_{2}n_{\bar{X}_{2}\bar{Y}_{2}})$}\\
%\vspace{0.1cm} 
{\footnotesize $~~~~~~~~~~~~~\Rightarrow~~m(X_{1}~ \rightarrow~ Y_{1}) = m(X_{2}~ \rightarrow~ Y_{2})]~and$}\\

{\footnotesize $[(n_{X_{1}\bar{Y}_{1}} = k_{1}n_{X_{2}\bar{Y}_{2}} ~ and~n_{\bar{X}_{1}\bar{Y}_{1}}= k_{1}n_{\bar{X}_{2}\bar{Y}_{2}}~and ~n_{X_{1}Y_{1}} = k_{2}n_{X_{2}Y_{2}} ~ and~n_{\bar{X}_{1}Y_{1}}= k_{2}n_{\bar{X}_{2}Y_{2}})$}\\
%\vspace{0.1cm} 
{\footnotesize $~~~~~~~~~~~~~\Rightarrow~~m(X_{1}~ \rightarrow~ Y_{1}) = m(X_{2}~ \rightarrow~ Y_{2})]$}\\

}}
%\end{tabular}
\end{center}

\vspace{0.3cm}

It is important to note that the formalization of this property by \cite{Tan02} with the help of the matrix is more compact than what we present, but in this article we are looking for the same formalization for all the properties.

\vspace{0.4cm}
{\bf \textcolor{blue}{Property 14} : Desired Relationship between the rules $X~ \rightarrow ~Y$ and $\bar{X}~ \rightarrow ~Y$.}

\begin{center}
%\begin{tabular}{|l|}
\fcolorbox[RGB]{230,230,250}{255,239,213}{\parbox{\linewidth}{

{\footnotesize
$P_{14}(m)~=~ 0~~~ if~~\exists~X~ \rightarrow ~Y~/~m(\bar{X}~\rightarrow ~Y) \neq -m(X~\rightarrow ~Y)$}\\

{\footnotesize
$P_{14}(m)~=~ 1~~~ if~~\forall~X~ \rightarrow ~Y~~~m(\bar{X}~\rightarrow ~Y) = -m(X~\rightarrow ~Y)$}\\
}}
%\end{tabular}
\end{center}

\vspace{0.4cm}
{\bf \textcolor{blue}{Property 15}: Desired Relationship between the antinomic rules $X~ \rightarrow ~Y$ and $X~ \rightarrow ~\bar{Y}$.}

\begin{center}
%\begin{tabular}{|l|}
\fcolorbox[RGB]{230,230,250}{245,245,245}{\parbox{\linewidth}{

{\footnotesize
$P_{15}(m)~=~ 0~~~ if~~\exists~X~ \rightarrow ~Y~/~m(X~ \rightarrow ~\bar{Y}) \neq -m(X~\rightarrow ~Y)$}\\

{\footnotesize
$P_{15}(m)~=~ 1~~~ if~~\forall~X~ \rightarrow ~Y~~~m(X~ \rightarrow ~\bar{Y}) = -m(X~\rightarrow ~Y)$}\\
}}
%\end{tabular}
\end{center}

\vspace{0.4cm}
{\bf \textcolor{blue}{Property 16}: Desired relationship between the rules $X~ \rightarrow ~Y$ and $\bar{X}~ \rightarrow ~\bar{Y}$.}

\begin{center}
%\begin{tabular}{|l|}
\fcolorbox[RGB]{230,230,250}{255,239,213}{\parbox{\linewidth}{
{\footnotesize
$P_{16}(m)~=~ 0~~~ if~~\exists~X~ \rightarrow ~Y~/~m(\bar{X}~\rightarrow ~\bar{Y}) \neq m(X~\rightarrow ~Y)$}\\

{\footnotesize
$P_{16}(m)~=~ 1~~~ if~~\forall~X~ \rightarrow ~Y~~~m(\bar{X}~\rightarrow ~\bar{Y}) = m(X~\rightarrow ~Y)$}\\
}}
%\end{tabular}
\end{center}

\vspace{0.4cm}
{\bf \textcolor{blue}{Property 17}: Premise size is fixed or random.}

\begin{center}
%\begin{tabular}{|l|}
%\hline
\fcolorbox[RGB]{230,230,250}{245,245,245}{\parbox{\linewidth}{
{\footnotesize
$~~P_{17}(m)~=~ 0~~(fixed~size)~~~~if~~m~isn't~established~on~a~probabilistic ~model$}\\

{\footnotesize
$P_{17}(m)~=~ 1~~(random~size)~if~~m~is~established~on~a~probabilistic~model$}\\
}}
%\end{tabular}
\end{center}

\vspace{0.4cm}
{\bf \textcolor{blue}{Property 18}: Descriptive or statistical measure.}

\begin{center}
%\begin{tabular}{|l|}
%\hline
\fcolorbox[RGB]{230,230,250}{255,239,213}{\parbox{\linewidth}{
{\footnotesize
$P_{18}(m)~=~ 0~~(descriptive~or~invariant)~ if~~\forall~k \in N^{*},~\forall~ X_{1}~ \rightarrow ~Y_{1},~\forall~ X_{2}~ \rightarrow ~Y_{2},$}\\
{\footnotesize $ (n_{X_{1}Y_{1}} = k n_{X_{2}Y_{2}}~~and~~n_{X_{1}\bar{Y}_{1}}= k n_{X_{2}\bar{Y}_{2}}~~and~~n_{\bar{X}_{1}Y_{1}} = k n_{\bar{X}_{2}Y_{2}} ~ ~and~~n_{\bar{X}_{1}\bar{Y}_{1}}= k n_{\bar{X}_{2}\bar{Y}_{2}})$}\\
{\footnotesize $~~~~~~~~~~~~~\Rightarrow~~m(X_{1}~ \rightarrow~ Y_{1}) = m(X_{2}~ \rightarrow~ Y_{2})$}\\
%\vspace{0.1cm} 
{\footnotesize
$P_{18}(m)~=~ 1~~(statistical)~ if~~\exists~k \in N^{*},~\exists~ X_{1}~ \rightarrow ~Y_{1},~\exists~ X_{2}~ \rightarrow ~Y_{2} /~(n_{X_{1}Y_{1}} = k n_{X_{2}Y_{2}}$}\\
{\footnotesize $and~~n_{X_{1}\bar{Y}_{1}}= k n_{X_{2}\bar{Y}_{2}}~~and~~n_{\bar{X}_{1}Y_{1}} = k n_{\bar{X}_{2}Y_{2}} ~ ~and~~n_{\bar{X}_{1}\bar{Y}_{1}}= k n_{\bar{X}_{2}\bar{Y}_{2}})$}\\
{\footnotesize $~~~~~~~~~~~~~and~~m(X_{1}~ \rightarrow~ Y_{1}) \neq m(X_{2}~ \rightarrow~ Y_{2})$}\\
}}
%\end{tabular}
\end{center}

\vspace{0.4cm}
{\bf \textcolor{blue}{Property 19}: Discriminant measure.}

\begin{center}
%\begin{tabular}{|l|}
%\hline
\fcolorbox[RGB]{119,136,153}{245,245,245}{\parbox{\linewidth}{
{\footnotesize
$P_{19}(m)~=~ 0~~(non~discriminant)~ if~~\exists~\eta \in N^{*} /~\forall~n > \eta~\forall~ X_{1}~ \rightarrow ~Y_{1},~\forall~ X_{2}~ \rightarrow ~Y_{2}$}\\
{\footnotesize $ [P(Y_{1}/X_{1}) > P(Y_{1})~~and~~P(Y_{2}/X_{2}) > P(Y_{2})]~~~\Rightarrow~~m(X_{1}~ \rightarrow~ Y_{1}) \simeq m(X_{2}~ \rightarrow~ Y_{2} $}\\
%\vspace{0.1cm} 
{\footnotesize
$P_{19}(m)~=~ 1~~(statistical)~~~~~~~~~~~ if~~\forall~\eta \in N^{*}~\exists~n > \eta~\exists~ X_{1}~ \rightarrow ~Y_{1}~\exists~ X_{2}~ \rightarrow ~Y_{2} /$}\\
{\footnotesize $ [P(Y_{1}/X_{1}) > P(Y_{1})~~and~~P(Y_{2}/X_{2}) > P(Y_{2})]~~~and~~m(X_{1}~ \rightarrow~ Y_{1}) \neq m(X_{2}~ \rightarrow~ Y_{2}) $}\\
}}
%\hline
%\end{tabular}
\end{center}

After formalizing the properties, we will study them on the different obectives measures.

\section{Evaluation of properties on measures}
\label{sec:evalprop}

This section looks for different objective interestingness measures, the presence or absence of the properties identified in \textit{Section} \ref{subsec:prop} and formalized in \textit{Section} \ref{subsec:formprop}. This work will lead to the construction of a measure-property matrix. 
%which can be visualized in \textit{Table 1}.

We examined $69$ measures of which $46$ are from synthesis work \cite{PS91}, \cite{Tan02}, \cite{lallich04}, \cite{GengH07:crl_fwidm}, \cite{Vaillant06} and \cite{Feno07}. Nine measures described in \cite{Huynh06arqat} have also been studied. These measures include: \textit{\{Causal confidence, Causal confirmed confidence, Descriptive confirmed confidence, Causal confirmation, Descriptive confirmation, Dependency, Putative causal dependency, Pavillon and Causal support\}}.

\begin{sloppypar}
Finally, the remaining measures are the following: \textit{Czekanowski-Dice} \cite{Czek}, \textit{Fukuda} \cite{Fukuda96}, \textit{Ganascia} \cite{Ganascia87}, \textit{probabilistic index of deviation from equilibrium} \cite{Blanchard2}, \textit{probabilistic index of deviation from the entropic equilibrium} \cite{Blanchard2}, \textit{entropic intensity of implication} \cite{gras01}, \textit{likelihood link index} \cite{Lerman81}, \textit{Kappa} \cite{Cohen60}, \textit{Kulczynski} \cite{Kulcz}, \textit{MGK} \cite{Guillaume00}, \textit{Ochiai} \cite{Ochiai}, \textit{satisfaction} \cite{Lavrac99} and \textit{VT100} \cite{Morineau06}.

Through the study of these different measures, we detect the presence of measures having the same definition but different names and are as follows: 

\begin{center}

\begin{itemize}
\item \{$\varphi-coefficient$ or \textit{Correlation coefficient}\};
\item \{\textit{Cohen} or \textit{Kappa}\};
\item \{\textit{Centred confidence} or \textit{Added value} or \textit{Pavillon}\};
\item \{\textit{Descriptive-confirmed confidence} or \textit{Ganascia}\};
\item \{\textit{Cosine} or \textit{Ochiai}\};
\item \{\textit{Czekanowski-Dice} or \textit{F-measure}\};
\item \{\textit{Bayes factor} or \textit{Odd-multiplier}\};
\item \{\textit{Factor of certainty} or \textit{Satisfaction} or \textit{Loevinger}\};
\item \{\textit{Kulczynski} or \textit{Agreement and disagreement index}\};
\item \{\textit{Support} or  \textit{Russel and Rao index}\};
\item \{\textit{Accuracy} or \textit{Causal support}\}.
\end{itemize}

\end{center}
%%%Rajouter ce que nous avons mit dans ICDM

\end{sloppypar}

Therefore, if we keep only one measure from the ones listed above, we will be in the presence of $61$ measures. \textit{Table}~\ref{tab:list_meas} summarizes and groups them into two categories: symmetrical and asymmetrical measures. The definition of each index is available in {\em Appendix 1\/} in \textit{table}~\ref{tab:defmeasure}. The $61$ measures of the table are ordered alphabetically, the number of measures given in the table facilitates the search for its definition.
After presenting data on which we will achieve a classification, we now ensure that they can not be constrained by searching for groups of measures with identical behavior and if properties are not redundant.

Initially, we searched all measures whose values for each of the $19$ properties are identical. We found the following seven groups:
$G_{1}$ = {{\em correlation coefficient, novelty \/}}, $G_{2}$ = {{\em Causal confidence, Causal-confirm confidence, Negative reliability\/}}, $G_{3}$ = {{\em Cosine, Czekanowski-Dice \/}}, $G_{4}$ = {{\em Causal dependency, Leverage, Specificity\/}}, $G_{5}$ = {{\em Collective strength, Odds ratio\/}}, $G_{6}$ = {{\em Gini, Mutual information\/}} and $G_{7}$ = {{\em Jaccard, Kulczynski\/}}.

Following the detection of these seven groups of measures, we are now in the presence of a matrix of $52$ measures since we retain only one measure from each one.

By looking if properties are not redundant, we investigated whether a property had identical values with another property for each of the $52$ measures. We haven't found such relationship.

\section{Categorization}\label{sec:cat}

Actually, we are in the presence of a matrix of $52$ measures and $19$ properties, properties that are nominal qualitative variables. Nevertheless, it's not easy for data mining experts to choose the appropriate interestingness measure from a set of $52$ measures. Therefore, it is frequently necessary to identify groups of measures with similar properties to help the user capture the most suitable ones. The most commonly used technique for finding such relationships is cluster analysis \cite{FayyadPS96}, \cite{hartigan75}.

Clustering techniques are generally used in an unsupervised fashion. They are used to place data elements into several groups such that elements in the same group are close to each others and elements across groups are far from each others \cite{dudaHart1973}. However, there exist many efficient clustering algorithms in the data mining literature among which the well-known and used are \textit{k-means} clustering and Agglomerative Hierarchical Clustering (\textit{AHC}). Choosing one of those techniques is not an easy task, if each of them has advantages and limitations.

\subsection{\textit{K-means} technique}

\textit{K-means} clustering \cite{Mac67} is a commonly used method \cite{Bradley98scalingclustering}, \cite{Farnstrom:2000}, \cite{Roweis:1999} of cluster analysis which aims to automatically partition observations into \textit{k} groups of greatest possible distinction, where \textit{k} is provided as an input parameter. It is an iterative aggregation method which, wherever it starts from, converges on a solution. K-means has several advantages. It is simple and fast: with a large number of variables, it may be computationally faster than hierarchical clustering (when \textit{k} is small). In addition, any element may be assigned to a group during one iteration then change from group in the following iteration, which is not possible with \textit{AHC} for which assignment is irreversible. 

Despite these advantages, the fixed number of clusters that k-means clustering technique require to specify as an input, can make it difficult to predict the appropriate number of clusters \textit{k}. Then, an inappropriate choice of \textit{k} may yield to poor results. Another disadvantage to using this technique is the possibility of multiplying the starting locations of cluster centers, which yield to several solutions and multiple clusterings. The solution obtained is not necessarily the same for all starting points.

\subsection{\textit{AHC} technique}

In data mining, hierarchical clustering \cite{Ward:hgobf} is a one of the most frequently method of cluster analysis which seeks to build a hierarchy of clusters. Agglomerative hierarchical clustering \cite{Guha:1998}, \cite{Guha:2000}, \cite{Karypis:Chameleon}, \cite{King:1967}, \cite{Sneath:1973} is a "bottom-up" clustering method where each observation starts in its own cluster, and pairs of clusters are merged as one moves up the hierarchy. Hierarchical clustering solutions, which are in the form of trees called dendrograms, are of great interest for a number of application domains. Despite its proven utility, hierarchical clustering has many flaws: e.g., interpretation of the hierarchy is complex and often confusing; the use of different distance metrics for measuring distances between clusters may generate different results. Nevertheless, it is also essential to recognize the advantages of \textit{AHC}, if it can produce an ordering of the elements, which may be informative for data display. Smaller clusters are generated, which may be helpful for discovery.

The importance revealed by the agglomerative hierarchical clustering and \textit{k-means} clustering techniques, encourage us to apply both of them on our measure-property matrix in order to come out with a consensus.

To launch two versions of clustering algorithms, versions require binary variables, we perform a complete disjunctive encoding, which leads us to obtain $39$ binary variables. So we have finally a matrix of $52$ measures $\times$ $39$ binary variables.

After discussing the data and converted them to be able to apply the selected algorithms, we study the first clustering of measures obtained with a method of hierarchical cluster analysis.

\section{Classification obtained by AHC method}
\label{sec:clacah}

\begin{figure}
\centering
\includegraphics[scale=.9]{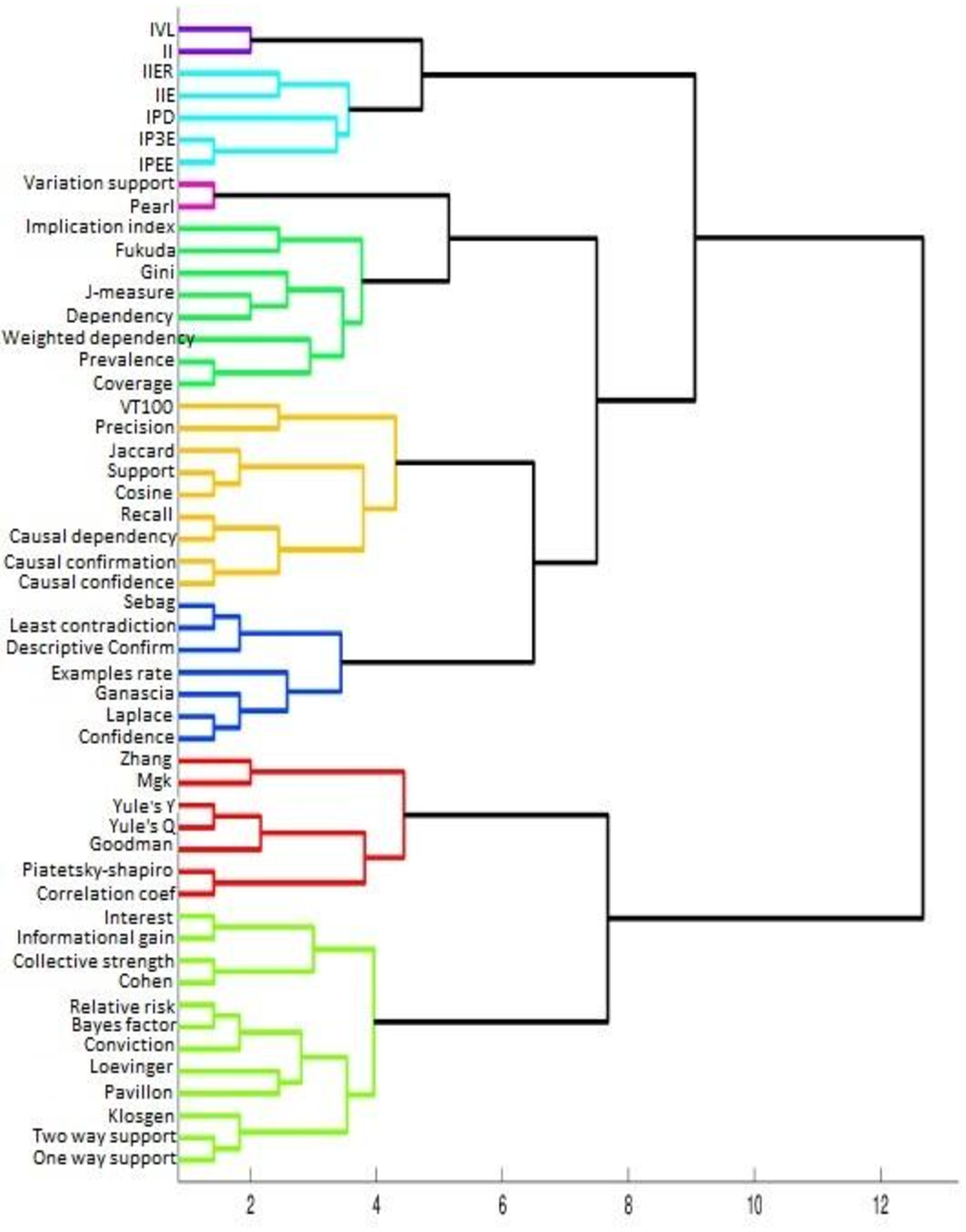}
\caption{Agglomerative hierarchical clustering using Ward criterion}
\label{fig:dendogram}
\end{figure}

We made an agglomerative hierarchical classification with {\em Matlab \/} software on these $52$ measures using Euclidean distance between pairs of measures then Ward distance for the aggregation phase. {\em Figure\/}~\ref{fig:dendogram} restitues this classification for Ward distance. As the loss of interclass inertia must be as small as possible, we cut the dendrogram at a level where branch height is high, corresponding to the dendogram colored branches.

\begin{table}
\centering
%\rowcolors{1}{green}{pink}
\begin{tabular}{|p{0,4cm}|p{2,9cm}|p{0,4cm}|p{2,9cm}|} \hline
\multicolumn{4}{|c|}{\textbf{\textcolor{red}{Symmetric measures}}} \\ \hline
%\multirow{3}{*}{Immediate} & RR & Round Robin \\ \cellcolor{blue}
 \textcolor{blue}{1}  & correlation coefficient & \textcolor{blue}{2} & Cohen or Kappa \\ \hline
\textcolor{blue}{11} & Cosinus or Ochiai & \textcolor{blue}{13} & Czekanowski \\ \hline
\textcolor{blue}{20} & Collective strength & \textcolor{blue}{22} & Informationnel gain \\ \hline
\textcolor{blue}{24} & Goodman  & \textcolor{blue}{33} & Likelihood index  \\ \hline
\textcolor{blue}{34}  & interest  & \textcolor{blue}{35} & Jaccard \\ \hline
\textcolor{blue}{38} &  Kulczynski & \textcolor{blue}{43}  & Novelty  \\ \hline
\textcolor{blue}{44} & Pearl  & \textcolor{blue}{45}  & Piatetsky-Shapiro \\ \hline 
\textcolor{blue}{46}  & Accuracy  & \textcolor{blue}{48} & Yule's Q \\ \hline
\textcolor{blue}{50}  & Odds ratio  & \textcolor{blue}{54} & Support  \\ \hline
\textcolor{blue}{56} &  One way support & \textcolor{blue}{58}  & VT100 \\ \hline
\textcolor{blue}{59}  & Support variation  & \textcolor{blue}{60} &  Yule's Y  \\ \hline 
\multicolumn{4}{|c|}{\textbf{\textcolor{red}{Asymmetric measures}}} \\ \hline
\textcolor{blue}{3}  & Confidence  & \textcolor{blue}{4} & Causal confidence \\ \hline 
\textcolor{blue}{5}  & Pavillon & \textcolor{blue}{6}  & Ganascia  \\ \hline
\textcolor{blue}{7}  & Causal-confirm confidence  & \textcolor{blue}{8}  &  Causal confirm \\ \hline
\textcolor{blue}{9}  & Descriptive confirm  &  \textcolor{blue}{10} & Conviction \\ \hline 
\textcolor{blue}{12}  & Coverage & \textcolor{blue}{14} & Dependency \\ \hline
\textcolor{blue}{15}  & Causal dependency  & \textcolor{blue}{16}  & Weighted dependency  \\ \hline
\textcolor{blue}{17} & Bayes factor  & \textcolor{blue}{18}  & Factor of certainty or Loevinger  \\ \hline
\textcolor{blue}{19} & Negative reliability & \textcolor{blue}{21} & Fukuda \\ \hline
\textcolor{blue}{23} & Gini & \textcolor{blue}{25} & Implication index  \\ \hline   
\textcolor{blue}{26} & Probabilistic intensity of deviation from equilibrium (IPEE)  & \textcolor{blue}{27}  &  Entropic probabilistic intensity of deviation from equilibrium (IP3E) \\ \hline
\textcolor{blue}{28}  & Probabilistic discriminant index (PDI) & \textcolor{blue}{29} & Mutual information \\ \hline  
\textcolor{blue}{30}  & Intensity of Implication (II)  & \textcolor{blue}{31} &  Entropic intensity of implication (EII) \\ \hline  
\textcolor{blue}{32} & Entropic intensity of revised implication (REII)  & \textcolor{blue}{36}  & J-measure \\ \hline
\textcolor{blue}{37} & Klosgen & \textcolor{blue}{39} & Laplace \\ \hline
\textcolor{blue}{40} & Leverage & \textcolor{blue}{41} &  MGK \\ \hline   
\textcolor{blue}{42} & Least contradiction  & \textcolor{blue}{47} & Prevalence  \\ \hline
\textcolor{blue}{49} & Recall & \textcolor{blue}{51} & Relative risk \\ \hline  
\textcolor{blue}{52} & Sebag-Schoenauer & \textcolor{blue}{53} & Specificity  \\ \hline     
\textcolor{blue}{55} & One way support & \textcolor{blue}{57} & Examples rate \\ \hline
\textcolor{blue}{61} &  Zhang & &  \\ \hline 

\end{tabular}
\caption{Studied measures}
\label{tab:list_meas}
\end{table}

We might also choose the Manhattan distance and we would obtain similar results because the matrix is essentially binary: $18$ of $19$ binary variables, and in this case, Manhattan distance is the squared Euclidean distance. Only one variable has {\em three\/} values: property \textcolor{blue}{$P_{12}$}.\\

This classification reveals the {\em 8\/} following groups of measures:

\begin{itemize}
\item \textcolor[RGB]{127, 0, 255}{{\em $Gc_{1}$ = \{Likelihood index link, Intensity of implication (II)\} \/}}
\item \textcolor[RGB]{127, 0, 255}{{\em $Gc_{2}$ = \{REII, EII, PDI, IP3E, IPEE\}\/}}
\item \textcolor[RGB]{127, 0, 255}{{\em $Gc_{3}$ = \{Two-way variation Support, Pearl\}\/}}
\item \textcolor[RGB]{127, 0, 255}{{\em $Gc_{4}$ = \{Implication index, Fukuda, Gini, J-measure, Dependency, Weighted dependency, Prevalence, Coverage\}\/}}
\item \textcolor[RGB]{127, 0, 255}{{\em $Gc_{5}$ = \{VT100, Accuracy, Jaccard, Support, Cosine, Recall, Causal dependency, Causal confirm, Causal confidence\}\/}}
\item \textcolor[RGB]{127, 0, 255}{{\em $Gc_{6}$ = \{Sebag, Least contradiction, Descriptive confirmation, Examples rate, Ganascia, Laplace, Confidence\}\/}}
\item \textcolor[RGB]{127, 0, 255}{{\em $Gc_{7}$ = \{Zhang, MGK, Yule's Y, Yule's Q, Goodman, Piatetsky-Shapiro, Correlation coefficient\}\/}}
\item \textcolor[RGB]{127, 0, 255}{{\em $Gc_{8}$ = \{Interest, Informational gain, Collective strength, Cohen, Relative risk, Bayesian factor, Conviction, Factor of certainty, Pavilion, Klosgen, Two-way support, One-way support \}\/}}
\end{itemize}

After making this initial measures classification, we will compare it with the classification revealed by the second technique of the {\em k\/}-means method afterwards we discuss the different results obtained in order to reach a consensus.

\section{Classification obtained by a version of {\em k\/}-means}
\label{sec:clakmfi}

We performed a partitioning method with {\em k\/}-means using {\em Matlab\/} software by retaining equally the Euclidean distance. We chose {\em eight\/} classes according to the results of the {\em AHC\/} and we obtained the following partitioning. While presenting these {\em eight\/} new classes obtained, we discuss the consistency of the results obtained with the first technique.

\begin{itemize}
\item \textcolor[RGB]{127, 0, 255}{{\em $Gp_{1}$ = \{Likelihood index link, Intensity of implication (II), REII\} \/}}\\
This group is very close to the group $Gc_{1}$ since we have \textcolor[RGB]{127, 0, 255}{$Gp_{1}$ = $Gc_{1} \cup \{REII\}$}.
\item \textcolor[RGB]{127, 0, 255}{{\em $Gp_{2}$ = \{EII, PDI, IP3E, IPEE\}\/}}\\
This group is very close to the group $Gc_{2}$ since we have \textcolor[RGB]{127, 0, 255}{$Gp_{2}$ = $Gc_{2} - \{REII\}$}. We have the following equality: \textcolor[RGB]{127, 0, 255}{$Gp_{1} \cup Gp_{2} = Gc_{1} \cup Gc_{2}$}, which shows some consistency in the obtained results since we are in the presence of all indices of the likelihood link family.
\item \textcolor[RGB]{127, 0, 255}{{\em $Gp_{3}$ = \{Two-way variation Support, Pearl, Implication index, Gini, J-measure, Dependency, Prevalence, Coverage\}\/}}\\
This group is close to the group $Gc_{4}$ since we have:\\
\textcolor[RGB]{127, 0, 255}{$Gp_{3}$ = $Gc_{3} \cup Gc_{4} \cup ~\{Fukuda, Weighted~dependency\}$}. It should be noted that $Gc_{3}$ group, which is composed by {\em Two-way variation Support\/} and {\em Pearl\/} measures, is the closest group to $Gc_{4}$ (see dendogram in {\em figure\/}~\ref{fig:dendogram}).
\item \textcolor[RGB]{127, 0, 255}{{\em $Gp_{4}$ = \{Accuracy, Jaccard, Support, Cosine, Recall, Causal dependency, Causal confirm, Causal confidence\}\/}}\\
This group is similar to $Gc_{5}$ group since we have:\\
\textcolor[RGB]{127, 0, 255}{$Gp_{4} = Gc_{5}~ \cup ~\{Fukuda, Weighted~dependency\} - \{VT100\}$}.
\item \textcolor[RGB]{127, 0, 255}{{\em $Gp_{5}$ = \{Sebag, Least contradiction, Descriptive confirmation, Fukuda\}\/}}\\
This group is identical to $Gc_{6}$ group.
\item \textcolor[RGB]{127, 0, 255}{{\em $Gp_{6}$ = \{Zhang, MGK, Yule's Y, Yule's Q\}\/}}\\
This group is similar to $Gc_{7}$ group since we have:
\textcolor[RGB]{127, 0, 255}{$Gc_{7} = Gp_{6}~ \cup ~\{Piatetsky-Shapiro,~Correlation~coefficient\}$}
\item \textcolor[RGB]{127, 0, 255}{{\em $Gp_{7}$ = \{Interest, Informational gain, Relative risk, Bayes factor, Conviction, Certainty factor, Pavilion, Klosgen, Two-way support, One-way support\}\/}}\\
The group $Gp_{7}$ is very close to $Gc_{8}$ group since we have $10$ of $12$ measures in common. We have the following equality: \textcolor[RGB]{127, 0, 255}{$Gc_{8} = Gp_{7}~ \cup ~$ {\em {Collective~strength, Cohen}\/}.
\item \textcolor[RGB]{127, 0, 255}{{\em $Gp_{8}$ = \{VT100, Piatetsky-Shapiro, Correlation coefficient, Collective strength, Cohen\}\/}}}\\
Unlike other groups $Gp_{i} (i = \{1, .., 7\})$, this group is not similar to any of the $Gc_{j} (j = \{1, .., 8\})$ groups, since these {\em five\/} measures are from $Gc_{5}$, $Gc_{7}$ and $Gc_{8}$ groups.
\end{itemize}

%Weighted dependency, Goodman, Examples rate, Ganascia, Laplace, Confidence , 
A consensus on the classification is presented in the following.

\section{Final classification}\label{sec:fc}

\begin{figure}
\centering
\includegraphics[scale=.6]{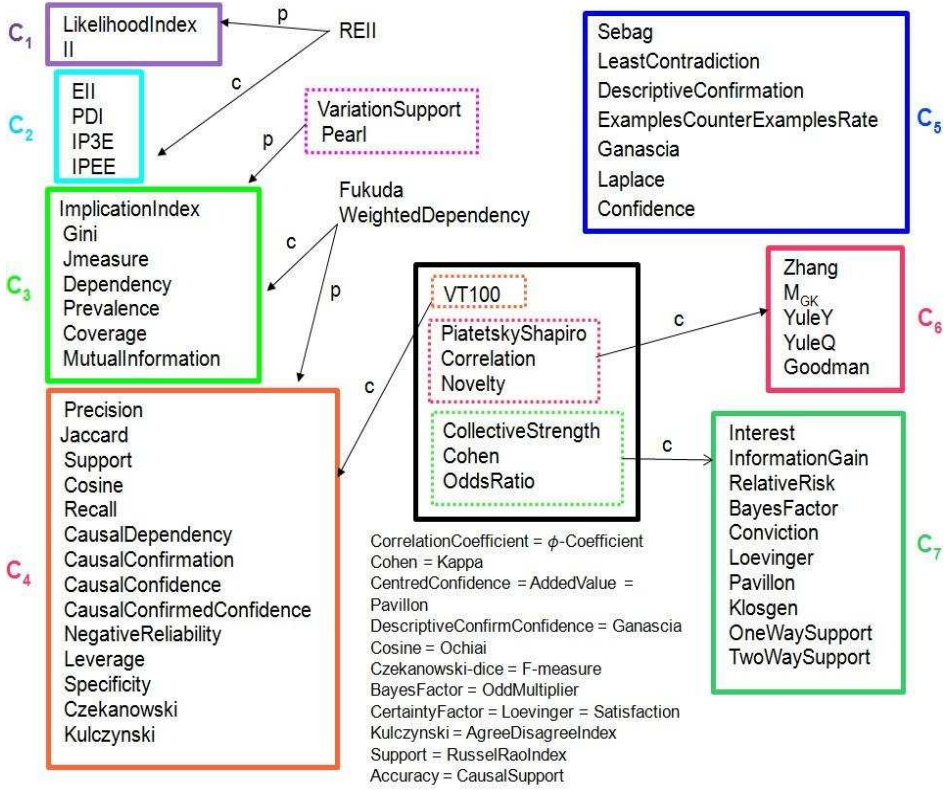}
\caption{Clusters of measures.}
\label{fig:ClassificationFormelle}
\end{figure}

After the discussion about the consistency of the results obtained by both techniques, we derive a consensus on the classification. {\em Figure\/}~\ref{fig:ClassificationFormelle} shows the consensus and restores the classes $C_{1}$ to $C_{7}$ of the common extracted measures to both techniques. We also include measures for which no consensus has been found and give, where it is possible, the two measures membership groups (or classes). We have labeled the arrows by "{\em c\/}" and "{\em p\/}" to indicate which technique gathered the measures in the pointed group ({\em c = hierarchical clustering or p = partitioning or non hierarchical clustering\/}). Finally, in the lower center of the figure, we recall the same measures but with different names.

Having summarized the results obtained ({\em Figure\/} \ref{fig:ClassificationFormelle}), we try in the next section to give a semantic to certain extracted classes and validate this classification with those released by \cite{Vaillant06}, \cite{Lesot:2010:OED}, \cite{ZighedAB11}.

\section{Clusters review and validation}
\label{sec:clreva}

It is not easy to give a semantic to each of the extracted classes by looking only the definitions of these measures. Two classes are yet easy to interpret, which are $C_{1}$ and $C_{2}$ classes where we find all the indices of the likelihood link index family \cite{Lerman70:apca}, the founder index. $C_{1}$ class has original indices: the likelihood link index and intensity of implication (II) \cite{Gras79}. We know that these two measures are very close since the likelihood link index searches if examples number (those who hold both the premise and conclusion) is significantly higher while Intensity of implication assesses whether the counter-examples number (those that satisfy the premise but does not verify the conclusion) is significantly low. 

For $C_{2}$ class, we find the Entropic implication intensity (EII \cite{gras01} and IP3E \cite{Blanchard2}) measures with the probabilistic index of deviation from equilibrium (IPEE \cite{Blanchard1}) and the probabilistic discriminant index  PDI \cite{LermanAze:QMDMbook07}. These measures are derived from a common idea: to assess the significance of a number (number of examples or counter-examples), combining for some measures (REII \cite{asmda05lvl}, EII, IP3E) with an entropic index so that the measure is discriminant in the case of large data. As for PDI, this index normalizes Intensity of implication in order that the latter be discriminant in the case of large data by evaluating a rule with respect to the set of valid rules.

To try to explain each of these classes $C_{i} (i = {1, .., 7})$, in {\em table\/} \ref{tab:evalmes}, we summarize all the properties satisfied by each of the seven classes. We add a symbol to the original matrix, the "$?$" character, which has the meaning "{\em unknown\/}" that is to say that measures of class $C_{i}$ take different values for the concerned property $P_{j} (j = {1, .., 19})$. In case where the property is a contradicted once, we show the majority property value. Then "$0?$" means that all the measures of the $C_{i}$ class except one measure, take the value "$0$" for the property $P_{j}$.

By summarizing all the properties satisfied by each of the seven classes in this table, we help the user in the selection of his measure(s) since he/she has only to read a much smaller matrix than the original. Moreover, if he/she wishes very different measures, his/her choice is also facilitated with the consultation of this table, help has been complemented by the dendrogram shown in the figure where a notion of proximity between measures appears. Finally, this classification can also help to choose measures that are too similar to avoid taking clues from the same class.

About finding semantic for each class, this synthetic table can be a support to an interpretation as we will illustrate it for $C_{4}$ and $C_{6}$ classes. We will therefore focus on these classes and try to give an interpretation. We start with the class $C_{6}$.

%\begin{table}
%\scriptsize
%\begin{center}
%\begin{tabular}{|p{0,25cm}|p{0,05cm}|p{0,05cm}|p{0,05cm}|p{0,05cm}|p{0,05cm}|p{0,05cm}|p{0,05cm}|p{0,05cm}|p{0,05cm}|p{0,05cm}|p{0,05cm}|p{0,05cm}|p{0,05cm}|p{0,05cm}|p{0,05cm}|p{0,05cm}|p{0,05cm}|p{0,05cm}|p{0,05cm}|}
%%\begin{tabular}{llllllllllllllllllll}
%\hline 
%Class & $P_{1}$ & $P_{2}$ & $P_{3}$ & $P_{4}$ & $P_{5}$ & $P_{6}$ & $P_{7}$ & $P_{8}$ & $P_{9}$ & $P_{10}$ & $P_{11}$ & $P_{12}$ & $P_{13}$ & $P_{14}$ & $P_{15}$ & $P_{16}$ & $P_{17}$ & $P_{18}$ & $P_{19}$ \\ 
%\hline 
%$C_{1}$ & ?	& 1	& 1	& 1	& 1	& 1	& 1	& 0	& 0	& 1	& 1	& 2	& ?	& 0	& 0	& 0	& 1	& 1	& 0	 \\ 
%\hline 
%$C_{2}$	& 1	& 1	& 1	& 1	& 0? & 1? & 0 & 0 & ? & 0 & 0 & 2 & 0? & 0 & 0	& 0	& 1	& 1	& 1? \\
%\hline 
%$C_{3}$ & 1	& ?	& 0? & 0 & 0 & 0 & ? & 0 & 0 & 0? & 0 & ? & 0 & 0 & 0 & ? & 0 & 0? &? \\ 
%\hline 
%$C_{4}$ & ?	& 1	& ?	& 1? & ? & 1? & 0 & ? & 0 & 0 & 0 & ? & 0 & 0 & 0 & 0? & 0 & 0 & 1 \\ 
%\hline 
%$C_{5}$ & 1	& 1	& ?	& 1	& 0	& 0	& 0	& ?	& 1	& 0	& 0	& ?	& 0	& 0	& ?	& 0	& 0	& 0	& 1? \\ 
%\hline 
%$C_{6}$ & ?	& 1	& 1	& 1	& 1	& 0	& 1	& 1	& 0	& 1	& 1	& ?	& ?	& ?	& 1	& ?	& 0	& 0	& 1 \\ 
%\hline 
%$C_{7}$ & ?	& 1	& ?	& ?	& 1	& 1? & 1 & 0? & 0 & 1 & 1 & ? & 0? & 0 & 0?	& 0	& 0	& 0	& 1 \\ 
%\hline 
%
%\end{tabular}
%\end{center}
%\caption{Characteristics of the seven detected classes} 
%\label{tab:evalmes}
%\end{table}

\begin{table}
%\scriptsize
\begin{center}
\begin{tabular}{|c|c|c|c|c|c|c|c|}
\hline 
\textcolor{blue}{Prop$\backslash$Clusters} & \textcolor{blue}{$C_{1}$} & \textcolor{blue}{$C_{2}$} & \textcolor{blue}{$C_{3}$} & \textcolor{blue}{$C_{4}$} & \textcolor{blue}{$C_{5}$} & \textcolor{blue}{$C_{6}$} & \textcolor{blue}{$C_{7}$} \\ 
\hline 
$P_{1}$ & ? & \textcolor{magenta}{1} & \textcolor{magenta}{1} & ? & \textcolor{magenta}{1} & ? & ? \\ 
\hline 
$P_{2}$ & \textcolor{magenta}{1} & \textcolor{magenta}{1} & ? & \textcolor{magenta}{1} & \textcolor{magenta}{1} & \textcolor{magenta}{1} & \textcolor{magenta}{1} \\ 
\hline 
$P_{3}$ & \textcolor{magenta}{1} & \textcolor{magenta}{1} & \textcolor{blue}{0?} & ? & ? & \textcolor{magenta}{1} & ? \\ 
\hline 
$P_{4}$ & \textcolor{magenta}{1} & \textcolor{magenta}{1} & \textcolor{blue}{0} & \textcolor{magenta}{1?} & \textcolor{magenta}{1} & \textcolor{magenta}{1} & ? \\ 
\hline 
$P_{5}$ & \textcolor{magenta}{1} & \textcolor{blue}{0?} & \textcolor{blue}{0} & ? & \textcolor{blue}{0} & \textcolor{magenta}{1} & \textcolor{magenta}{1} \\ 
\hline 
$P_{6}$ & \textcolor{magenta}{1} & \textcolor{magenta}{1?} & \textcolor{blue}{0} & \textcolor{magenta}{1?} & \textcolor{blue}{0} & \textcolor{blue}{0} & \textcolor{magenta}{1?} \\ 
\hline 
$P_{7}$ & \textcolor{magenta}{1} & \textcolor{blue}{0} & ? & \textcolor{blue}{0} & \textcolor{blue}{0} & \textcolor{magenta}{1} & \textcolor{magenta}{1} \\ 
\hline 
$P_{8}$ & \textcolor{blue}{0} & \textcolor{blue}{0} & \textcolor{blue}{0} & ? & ? &  \textcolor{magenta}{1} & \textcolor{blue}{0?} \\ 
\hline 
$P_{9}$ & \textcolor{blue}{0} & ? & \textcolor{blue}{0} & \textcolor{blue}{0} &  \textcolor{magenta}{1} & \textcolor{blue}{0} & \textcolor{blue}{0} \\ 
\hline 
$P_{10}$ & \textcolor{magenta}{1} & \textcolor{blue}{0} & \textcolor{blue}{0?} & \textcolor{blue}{0} & \textcolor{blue}{0} & \textcolor{magenta}{1} & \textcolor{magenta}{1} \\ 
\hline 
$P_{11}$ &\textcolor{magenta}{1} & \textcolor{blue}{0} & \textcolor{blue}{0} & \textcolor{blue}{0} & \textcolor{blue}{0}  & \textcolor{magenta}{1} & \textcolor{magenta}{1} \\ 
\hline 
$P_{12}$ & \textcolor{red}{2} & \textcolor{red}{2} & ? & ? & ?  & ? & ? \\ 
\hline 
$P_{13}$ & ? & \textcolor{blue}{0?} & \textcolor{blue}{0} & \textcolor{blue}{0} & \textcolor{blue}{0}  & ? & \textcolor{blue}{0?} \\ 
\hline 
$P_{14}$ & \textcolor{blue}{0} & \textcolor{blue}{0} & \textcolor{blue}{0} & \textcolor{blue}{0} & \textcolor{blue}{0}  & ? & \textcolor{blue}{0} \\ 
\hline 
$P_{15}$ & \textcolor{blue}{0} & \textcolor{blue}{0} & \textcolor{blue}{0} & \textcolor{blue}{0} & ? & 1 & \textcolor{blue}{0?} \\ 
\hline 
$P_{16}$ & \textcolor{blue}{0} & \textcolor{blue}{0} & ? & \textcolor{blue}{0?} & \textcolor{blue}{0} & ? & \textcolor{blue}{0} \\ 
\hline 
$P_{17}$ & \textcolor{magenta}{1} & \textcolor{magenta}{1} & \textcolor{blue}{0} & \textcolor{blue}{0} & \textcolor{blue}{0} & \textcolor{blue}{0} & \textcolor{blue}{0} \\ 
\hline 
$P_{18}$ & \textcolor{magenta}{1} & \textcolor{magenta}{1} & \textcolor{blue}{0?} & \textcolor{blue}{0} & \textcolor{blue}{0} & \textcolor{blue}{0} & \textcolor{blue}{0} \\ 
\hline 
$P_{19}$ & \textcolor{blue}{0} & \textcolor{magenta}{1?} & ? & \textcolor{magenta}{1} & \textcolor{magenta}{1?} & \textcolor{magenta}{1} & \textcolor{magenta}{1} \\ 
\hline 
\end{tabular} 
\end{center}
\caption{Characteristics of the seven detected classes} 
\label{tab:evalmes}
\end{table}

\subsection{$C_{6}$ class study}
\label{subsec:stdC6}

$C_{6}$ class is composed of five measures: Zhang \cite{Zhang}, MGK \cite{Guillaume00}, Y and Q of Yule \cite{Yule00:oas} and Goodman \cite{Tan02}. We know from the table that they satisfy the following properties:

\begin{itemize}
\item Non symmetry in the sense of conclusion negation (\textcolor{blue}{$P_{2} = 1$}),
\item Identical evaluation in the logical implication case (\textcolor{blue}{$P_{3} = 1$}),
\item Growth according to the number of examples (\textcolor{blue}{$P_{4} = 1$}),
\item Growth according to the data size (\textcolor{blue}{$P_{5} = 1$}),
\item Fixed value in the independence case (\textcolor{blue}{$P_{7} = 1$}),
\item Fixed value in the logical implication case (\textcolor{blue}{$P_{8} = 1$}),
\item Identifiable values when the realization of the premise increases the chances of occurrence of the conclusion (\textcolor{blue}{$P_{10} = 1$}),
\item Identifiable values when the realization of the premise reduces the chances of occurrence of the conclusion (\textcolor{blue}{$P_{11} = 1$}),
\item Opposed values for the antinomic rules $ X \rightarrow Y $ and $ X \rightarrow \bar{Y}  $ (\textcolor{blue}{$P_{15} = 1$}),
\item Discriminant in the case of large data (\textcolor{blue}{$P_{19} = 1$}).
\end{itemize}

Due to the set of  satisfied properties, we can give a first semantic for $C_{6}$ class. These measures are a standardized indices since they have a fixed values for the independence (\textcolor{blue}{$P_{7} = 1$}) and logical implication (\textcolor{blue}{$P_{8} = 1$}) case and the values taken by these indices to determine whether the rule is in the attractive (\textcolor{blue}{$P_{10} = 1$}) or in the repulsive area (\textcolor{blue}{$P_{11} = 1$}).

\begin{figure}
\centering
\makebox[0cm][r]{\includegraphics[width=0.45\linewidth]{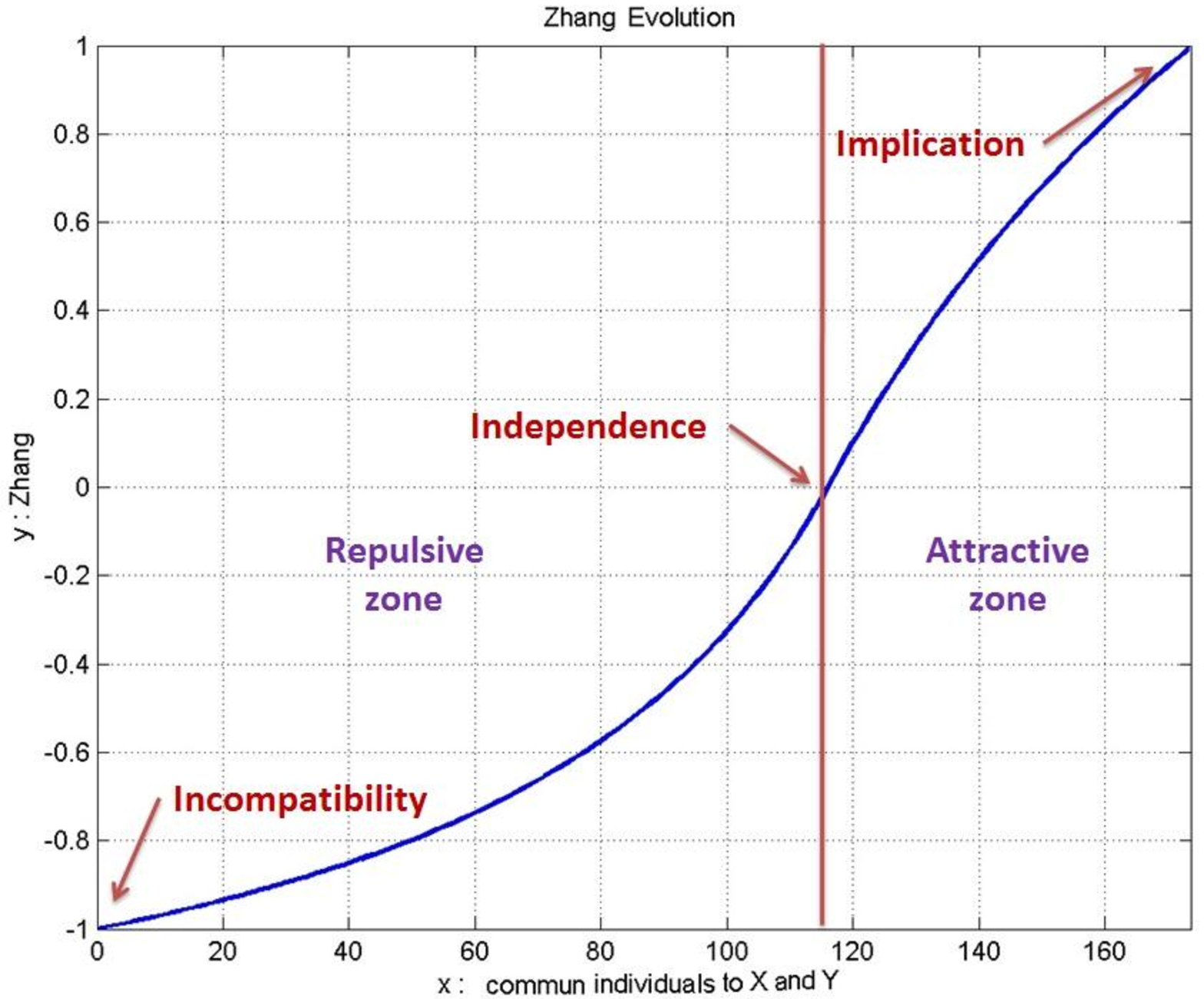}}
\hspace{0.7cm}
\makebox[0cm][l]{\vspace{0.5\linewidth}\includegraphics[width=0.43\linewidth]{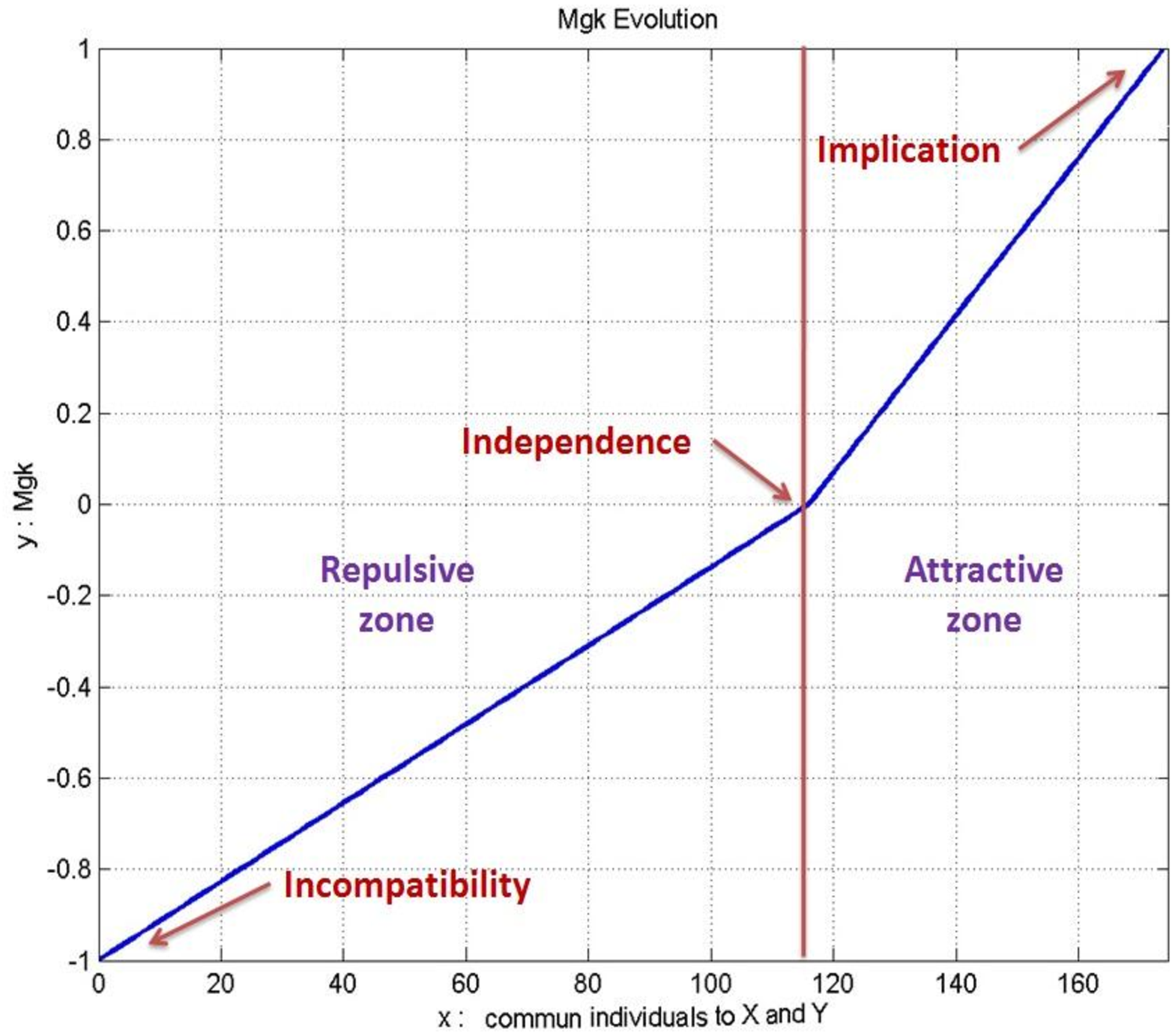}}\\
\vspace{0.35cm}
\makebox[0cm][r]{\includegraphics[width=0.44\linewidth]{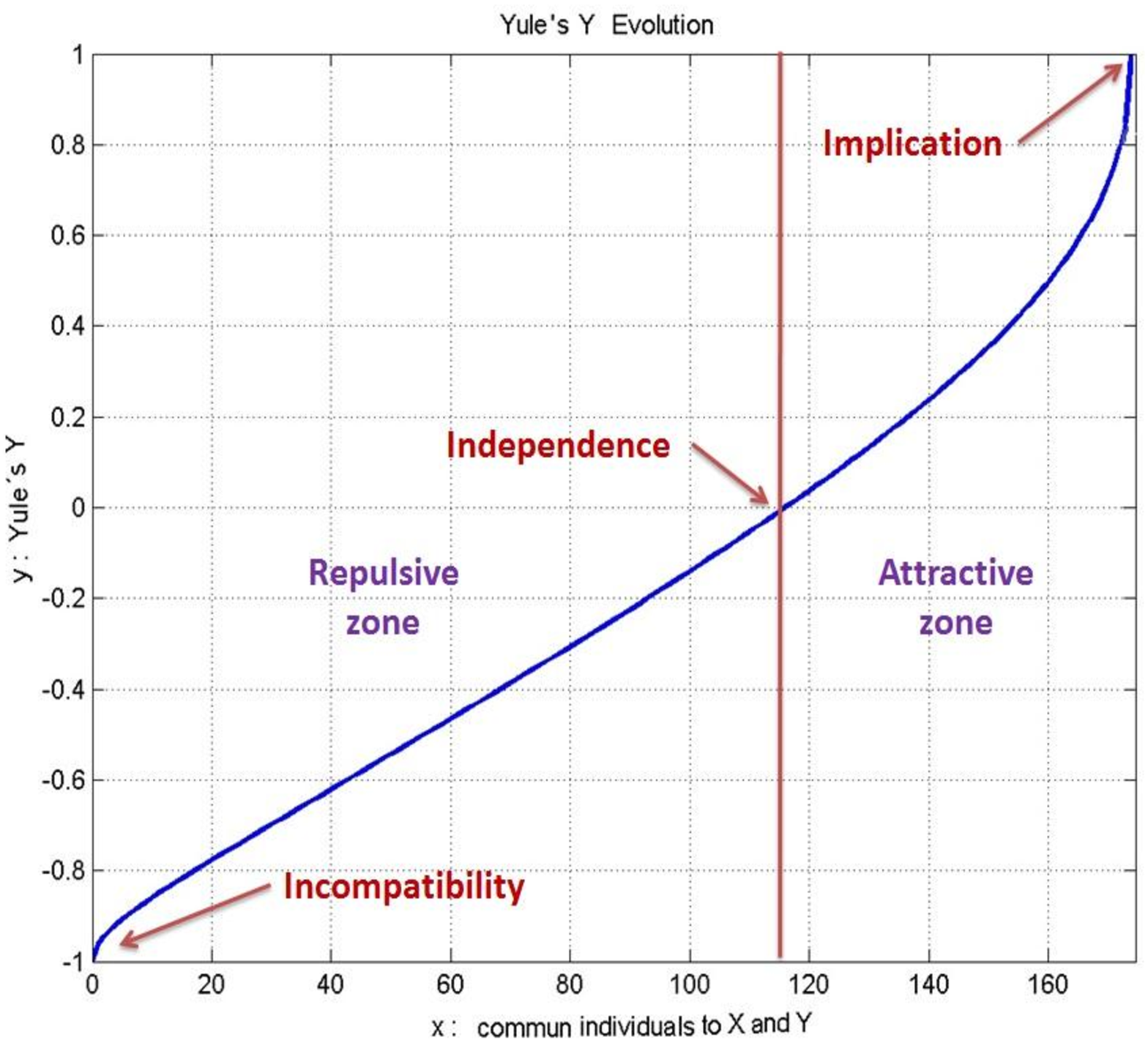}}
\hspace{0.7cm}
\makebox[0cm][l]{\vspace{0.5\linewidth}\includegraphics[width=0.45\linewidth]{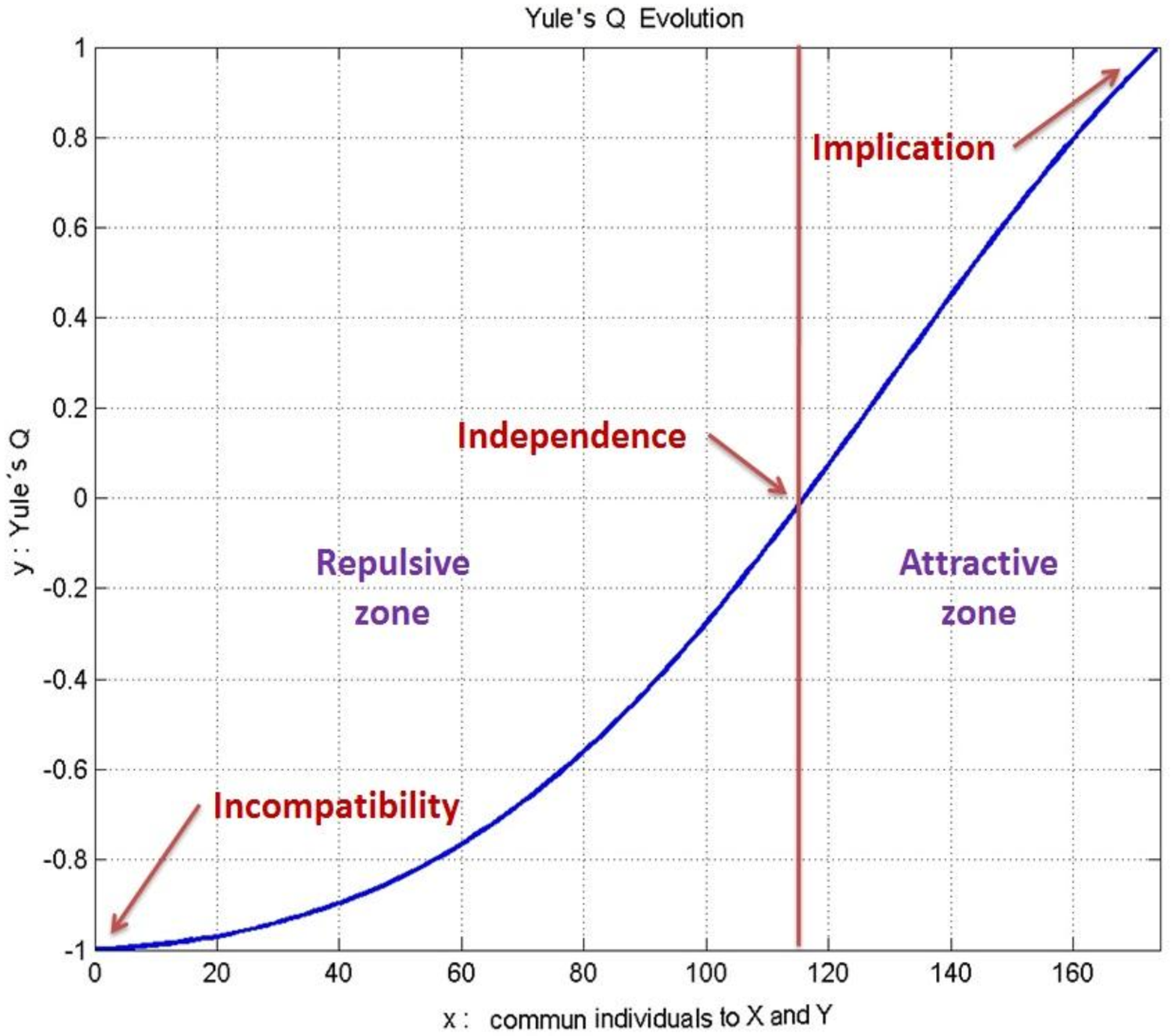}}\\
\vspace{0.35cm}
\includegraphics[width=0.45\linewidth]{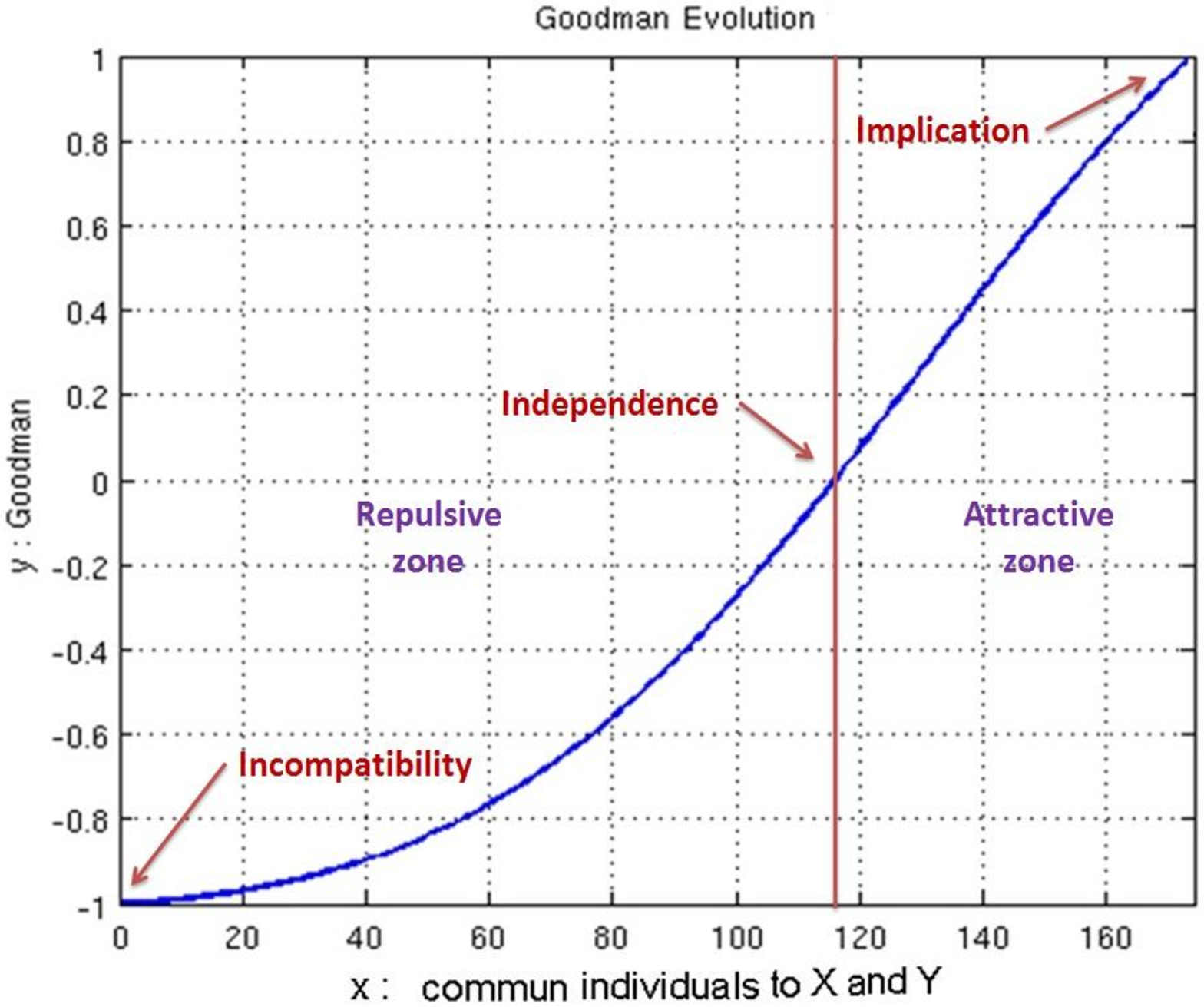}
\vspace{0.3cm}
\caption{Evolution of the five measures of $C_{6}$ class according to the number of examples.}
\label{fig:C6Measures}
\end{figure}

{\em Figure\/}~\ref{fig:C6Measures} enables to verify the first semantic given to these indexes. We traced the evolution of the five measures when the number of examples increases starting then from the incompatibility state (no individual checks both the premise and the conclusion or also $n_{XY} = 0$, with $n_{XY}$ the number of individuals verifying both the premise $X$ and conclusion $Y$) to the logical implication (The set of individuals verifying the premise is included in the set of individuals satisfying the conclusion or also $n_{XY} = n_{X}$ with $n_{X}$ the number of individuals satisfying the premise $X$). As well, we have shown in {\em figure\/}~\ref{fig:C6Measures} the three characteristic states of a rule: the incompatibility, independence and logical implication in addition to the attraction and repulsion areas. The whole premise size used to carry out these curves is $174$, the overall conclusion size is $400$ and finally the dataset size is $600$ ($n_{X} = 174$, $n_{Y} = 400$ and $n = 600$). We could have chosen different sizes for these different sets and would have obtained similar curves observed with the following constraint: $n_{X} \leq n_{Y} \leq n$.

{\em Figure\/}~\ref{fig:C6Measures} allows us to refine the semantic given to this class $C_{6}$. These are standardized measures with values between $-1$ and $1$ with fixed values equal to $-1$, $0$ and $1$ respectively for the incompatibility, independence and logical implication. Moreover, they don't have only identifiable values in the attraction and repulsion area, but these values are between $0$ and $1$ in the attraction area and between $-1$ and $0$ in the repulsion area. Finally, the measure sign provides information about the area belonging to the rule. We can deduce that these measures assess a certain distance according to the independence: distance between the independence and the logical implication in case of positive values and a distance between independence and incompatibility in case of negative values.

%\begin{table}
%\centering
%\label{tab:evalmes}
%\caption{Evaluation of properties on the measures of Class 6.}
%%\rowcolors{1}{green}{pink}
%%\begin{tabular}{|p{0,4cm}|p{2,9cm}|p{0,4cm}|p{2,9cm}|} \hline
%
%\begin{tabular}{|c|c|c|c|c|c|c|c|c|c|c|c|c|c|c|c|c|c|c|c|}
%\hline 
%Classes & $P_{1}$ & $P_{2}$ & $P_{3}$ & $P_{4}$ & $P_{5}$ & $P_{6}$ & $P_{7}$ & $P_{8}$ & $P_{9}$ & $P_{10}$ & $P_{11}$ & $P_{12}$ & $P_{13}$ & $P_{14}$ & $P_{15}$ & $P_{16}$ & $P_{17}$ & $P_{18}$ & $P_{19}$ \\ 
%\hline    
%Zhang	& 1	& 1	& 1	& 1	& 1	& 0	& 1	& 1	& 0	& 1	& 1	& 2	& 1	& 0	& 1	& 0	& 0	& 0	& 1	\\  
%\hline  
%MGK	    & 1	& 1	& 1	& 1	& 1	& 0	& 1	& 1	& 0	& 1	& 1	& 1	& 0	& 0	& 1	& 0	& 0	& 0	& 1	\\ 
%\hline  
%Yule'Y	& 0	& 1	& 1	& 1	& 1	& 0	& 1	& 1	& 0	& 1	& 1	& 0	& 1	& 1	& 1	& 1	& 0	& 0	& 1	\\
%\hline   
%Yule'Q	& 0	& 1	& 1	& 1	& 1	& 0	& 1	& 1	& 0	& 1	& 1	& 2	& 1	& 1	& 1	& 1	& 0	& 0	& 1	\\
%\hline    
%Goodman	& 0	& 1	& 1	& 1	& 1	& 0	& 1	& 1	& 0	& 1	& 1	& 1	& 0	& 1	& 1	& 1	& 0	& 0	& 1	\\
%\hline    
%Class 6 & ?	& 1	& 1	& 1	& 1	& 0	& 1	& 1	& 0	& 1	& 1	& ?	& ?	& ?	& 1	& ?	& 0	& 0	& 1	\\
%\hline
%\end{tabular} 
%\end{table} 

\begin{table}
\centering
\begin{tabular}{|c|c|c|c|c|c|c|}
\hline 
\textcolor{blue}{Property} & \textcolor{blue}{Zhang} & \textcolor{blue}{MGK} & \textcolor{blue}{Yule'Y} & \textcolor{blue}{Yule'Q} & \textcolor{blue}{Good} & \textcolor{blue}{C6} \\ 
\hline 
$P_{1}$ & \textcolor{magenta}{1} & \textcolor{magenta}{1} & \textcolor{blue}{0} & \textcolor{blue}{0} & \textcolor{blue}{0} & ? \\ 
\hline 
$P_{2}$ & \textcolor{magenta}{1} & \textcolor{magenta}{1} & \textcolor{magenta}{1} & \textcolor{magenta}{1} & \textcolor{magenta}{1} & \textcolor{magenta}{1} \\ 
\hline 
$P_{3}$ & \textcolor{magenta}{1} & \textcolor{magenta}{1} & \textcolor{magenta}{1} & \textcolor{magenta}{1} & \textcolor{magenta}{1} & \textcolor{magenta}{1} \\ 
\hline 
$P_{4}$ & \textcolor{magenta}{1} & \textcolor{magenta}{1} & \textcolor{magenta}{1} & \textcolor{magenta}{1} & \textcolor{magenta}{1} & \textcolor{magenta}{1} \\ 
\hline 
$P_{5}$ & \textcolor{magenta}{1} & \textcolor[RGB]{127, 0, 255}{1} & \textcolor{magenta}{1} & \textcolor{magenta}{1} & \textcolor{magenta}{1} & \textcolor{magenta}{1} \\ 
\hline 
$P_{6}$ & \textcolor{blue}{0} & \textcolor{blue}{0} & \textcolor{blue}{0} & \textcolor{blue}{0} & \textcolor{blue}{0} & \textcolor{blue}{0} \\ 
\hline 
$P_{7}$ & \textcolor{magenta}{1} & \textcolor{magenta}{1} & \textcolor{magenta}{1} & \textcolor{magenta}{1} & \textcolor{magenta}{1} & \textcolor{magenta}{1} \\ 
\hline 
$P_{8}$ & \textcolor{magenta}{1} & \textcolor{magenta}{1} & \textcolor{magenta}{1} & \textcolor{magenta}{1} & \textcolor{magenta}{1} & \textcolor{magenta}{1} \\ 
\hline 
$P_{9}$ & \textcolor{blue}{0} & \textcolor{blue}{0} & \textcolor{blue}{0} & \textcolor{blue}{0} & \textcolor{blue}{0} & \textcolor{blue}{0} \\ 
\hline 
$P_{10}$ & \textcolor{magenta}{1} & \textcolor{magenta}{1} & \textcolor{magenta}{1} & \textcolor{magenta}{1} & \textcolor{magenta}{1} & \textcolor{magenta}{1} \\ 
\hline 
$P_{11}$ & \textcolor{magenta}{1} & \textcolor{magenta}{1} & \textcolor{magenta}{1} & \textcolor{magenta}{1} & \textcolor{magenta}{1} & \textcolor{magenta}{1} \\ 
\hline 
$P_{12}$ & \textcolor{red}{2} & \textcolor{magenta}{1} & \textcolor{blue}{0} &  \textcolor{red}{2} & \textcolor{magenta}{1} & ? \\ 
\hline 
$P_{13}$ & \textcolor{magenta}{1} & \textcolor{blue}{0} & \textcolor{magenta}{1} & \textcolor{magenta}{1} & \textcolor{blue}{0} & ? \\ 
\hline 
$P_{14}$ & \textcolor{blue}{0} & \textcolor{blue}{0} & \textcolor{magenta}{1} & \textcolor{magenta}{1} & \textcolor{magenta}{1} & ? \\ 
\hline 
$P_{15}$ & \textcolor{magenta}{1} & \textcolor{magenta}{1} & \textcolor{magenta}{1} & \textcolor{magenta}{1} & \textcolor{magenta}{1} & \textcolor{magenta}{1} \\ 
\hline 
$P_{16}$ & \textcolor{blue}{0} & \textcolor{blue}{0} & \textcolor{magenta}{1} & \textcolor{magenta}{1} & \textcolor{magenta}{1} & ? \\ 
\hline 
$P_{17}$ & \textcolor{blue}{0} & \textcolor{blue}{0} & \textcolor{blue}{0} & \textcolor{blue}{0} & \textcolor{blue}{0} & \textcolor{blue}{0} \\ 
\hline 
$P_{18}$ & \textcolor{blue}{0} & \textcolor{blue}{0} & \textcolor{blue}{0} & \textcolor{blue}{0} & \textcolor{blue}{0} & \textcolor{blue}{0} \\ 
\hline 
$P_{19}$ & \textcolor{magenta}{1} & \textcolor{magenta}{1} & \textcolor{magenta}{1} & \textcolor{magenta}{1} & \textcolor{magenta}{1} & \textcolor{magenta}{1} \\ 
\hline 
\end{tabular} 
\caption{Evaluation of properties on the measures of Class 6.}
\label{tab:evalmes6}
\end{table}

When we look at the figure showing the hierarchical clustering technique, we have a greater proximity between the indices Y, Q Yule and Goodman, and also higher proximity between Zhang and MGK. Discrepancies highlighted in the table, that is to say where we find the symbol "$?$" for the studied properties, we can learn about these two proximities more pronounced between the measures. Table \ref{tab:evalmes6} details the various properties satisfied by the five measures in this group and remember the general characteristics of this class. The first property where this symbol appears and which enables to explain these two proximities is the symmetry of measures (\textcolor{blue}{$ P_{1} $}). Y, Q Yule and Goodman are symmetric measures (similar assessment of the symmetrical rules $X \rightarrow Y$ and $Y \rightarrow X$: $ P_{1} = 0$) while Zhang and MGK are not symmetric measures (different evaluation of the symmetric rules $X \rightarrow Y$ and $Y \rightarrow X$: $ P_{1} = 1$).

Properties \textcolor{blue}{$ P_{14} $} (opposed values or not for the rules $X \rightarrow Y$ and $\bar{X} \rightarrow Y$) and \textcolor{blue}{$ P_{16} $} (identical values for the rules $X \rightarrow Y$ and $\bar{X} \rightarrow \bar{Y}$ or not) also help to explain these two proximities. Indices Y, Q and Goodman have opposite values for the rules $X \rightarrow Y$ and $\bar{X} \rightarrow Y$ and identical values for the rules $X \rightarrow Y$ and $\bar{X} \rightarrow \bar{Y}$. The measures Zhang and MGK verify the negation of the two latter properties.\\

We will now make a study of class $C_{4}$.

\subsection{Study of the $C_{4}$ class}
\label{subsec:stdC4}

Class $C_{4}$ contains the following indexes: \textcolor[RGB]{127, 0, 255}{{\em Accuracy\/}} \cite{Tan02}, \textcolor[RGB]{127, 0, 255}{{\em Jaccard\/}} \cite{jaccard}, \textcolor[RGB]{127, 0, 255}{{\em Support\/}} \cite{Russel40:hasalsm}, \textcolor[RGB]{127, 0, 255}{{\em Cosine\/}} \cite{Ochiai}, \textcolor[RGB]{127, 0, 255}{{\em Recall\/}} \cite{Lavrac99}, \textcolor[RGB]{127, 0, 255}{{\em Causal dependency\/}} \cite{Tan02}, \textcolor[RGB]{127, 0, 255}{{\em Causal confidence\/}} \cite{Kodratoff:2001:CML}, \textcolor[RGB]{127, 0, 255}{{\em Causal-confirm confidence\/}} \cite{Kodratoff:2001:CML}, \textcolor[RGB]{127, 0, 255}{{\em Negative reliability\/}} \cite{Lavrac99}, \textcolor[RGB]{127, 0, 255}{{\em Leverage\/}} \cite{PS91}, \textcolor[RGB]{127, 0, 255}{{\em Specificity\/}} \cite{Tan02}, \textcolor[RGB]{127, 0, 255}{{\em Czekanowski-Dice\/}} \cite{Czek} and \textcolor[RGB]{127, 0, 255}{{\em Kulczynski\/}} \cite{Kulcz}.\\

From {\em table\/}~\ref{tab:evalmes}, these $14$ measures satisfy the $12$ following properties:

\begin{itemize}
\item Non symmetry in the sense of conclusion negation (\textcolor{blue}{$P_{2} = 1$}),
\item Discriminant in the case of large data (\textcolor{blue}{$P_{19} = 1$}),
\item Non Fixed value in the independence case (\textcolor{blue}{$P_{7} = 0$}) and equilibrium (\textcolor{blue}{$P_{9} = 0$}),
\item Unidentifiable values in the case of attraction (\textcolor{blue}{$P_{10} = 0$}) and repulsion (\textcolor{blue}{$P_{11} = 0$}),
\item Non-invariant in the case of expansion of certain numbers (\textcolor{blue}{$P_{13} = 0$}),
\item Two relations between the different negative rules are not present (\textcolor{blue}{$P_{14} = 0$}) (\textcolor{blue}{$P_{15} = 0$}),
\item Not based on a probabilistic model (\textcolor{blue}{$P_{17} = 0$}),
\item Descriptive measures (\textcolor{blue}{$P_{18} = 0$}).\\
Let us study now the properties satisfied by almost all the measures except one:
\item Growth according to the number of examples (\textcolor{blue}{$P_{4} = 1$}) with the exception of the {\em \textcolor[RGB]{127, 0, 255}{Support}\/},
\item Growth according to the size of the conclusion (\textcolor{blue}{$P_{6} = 1$}) with the exception of the {\em \textcolor[RGB]{127, 0, 255}{Support}\/},
\item Measures do not equalize the rules $X \rightarrow Y$ and $\bar{X} \rightarrow \bar{Y}$ (\textcolor{blue}{$P_{16} = 0$}) with the exception of {\em \textcolor[RGB]{127, 0, 255}{Accuracy}\/}.
\end{itemize}

Given the relatively large number of the measures present in this class (the class whose cardinality is greater), it is difficult to find a semantic as precise as for the previous class $C_{6}$. However, we can give one to a smaller set of measures: {\em \textcolor[RGB]{127, 0, 255}{Jaccard}, \textcolor[RGB]{127, 0, 255}{Support}, \textcolor[RGB]{127, 0, 255}{Cosine}, \textcolor[RGB]{127, 0, 255}{Czekanowski-Dice}, \textcolor[RGB]{127, 0, 255}{Kulczynski} and \textcolor[RGB]{127, 0, 255}{Recall}\/}. These measures are function $P(XY)$ and symmetrical (with the exception of the Recall). We recall the expressions of these {\em six\/} measures:

\begin{itemize}
\item $ \textcolor[RGB]{127, 0, 255}{Jaccard} :  \frac{P(XY)} {P(X) + P(Y) - P(XY)} = \frac{P(XY)} {P(X\bar{Y}) + P(Y)} $
\item $ \textcolor[RGB]{127, 0, 255}{Support} : P(XY) $
\item $ \textcolor[RGB]{127, 0, 255}{Cosine} : \frac{P(XY)}{\sqrt{P(X)P(Y)}} $
\item $ \textcolor[RGB]{127, 0, 255}{Czekanowski-Dice} :  \frac{2 P(XY)}{P(X) + P(Y)} $
\item $ \textcolor[RGB]{127, 0, 255}{Kulczynski} : \frac {P(XY)} {P(X\bar{Y}) + P(\bar{X}Y)} $
\item $ \textcolor[RGB]{127, 0, 255}{Recall} : \frac {P(XY)} {P(Y)} $
\end{itemize}

We can then deduce that these measures will have a fixed value equal to $0$ in the case of incompatibility ($P(XY) = 0$). We also understand the non growth that is founded according to the dataset size ($P_{5} = 0$) at the sight of these different formulas as shown in {\em table\/}~\ref{tab:evalc4} which reproduces the satisfied properties by these {\em six\/} measures. We have an invariance of these measures ({\em except for the \textcolor[RGB]{127, 0, 255}{Support}\/}) depending on the size {\em n\/} of the dataset since it amounts to increase the probability $P(\bar{XY})$. As to the {\em \textcolor[RGB]{127, 0, 255}{Support}\/}, it is decreasing according to the size {\em n\/} of the whole data.

%\begin{table}
%\centering
%\label{tab:evalmes}
%\caption{Evaluation of properties on a subset of measures of Class 4.}
%\rowcolors{1}{green}{pink}
%\begin{tabular}{|p{0,4cm}|p{2,9cm}|p{0,4cm}|p{2,9cm}|} \hline

%\begin{tabular}{|c|c|c|c|c|c|c|c|c|c|c|c|c|c|c|c|c|c|c|c|}
%\hline 
%Classes & $P_{1}$ & $P_{2}$ & $P_{3}$ & $P_{4}$ & $P_{5}$ & $P_{6}$ & $P_{7}$ & $P_{8}$ & $P_{9}$ & $P_{10}$ & $P_{11}$ & $P_{12}$ & $P_{13}$ & $P_{14}$ & $P_{15}$ & $P_{16}$ & $P_{17}$ & $P_{18}$ & $P_{19}$ \\ 
%\hline    
%Jaccard	    & 0	& 1	& 0	& 1	& 0	& 1	& 0	& 0	& 0	& 0	& 0	& 0	& 0	& 0	& 0	& 0	& 0	& 0	& 1	\\  
%\hline  
%support	    & 0	& 1	& 0	& 0	& 0	& 0	& 0	& 0	& 0	& 0	& 0	& 1	& 0	& 0	& 0	& 0	& 0	& 0	& 1	\\ 
%\hline  
%cosinus	    & 0	& 1	& 0	& 1	& 0	& 1	& 0	& 0	& 0	& 0	& 0	& 1	& 0	& 0	& 0	& 0	& 0	& 0	& 1 \\
%\hline   
%recall	    & 1	& 1	& 0	& 1	& 0	& 1	& 0	& 0	& 0	& 0	& 0	& 1	& 0	& 0	& 0	& 0	& 0	& 0	& 1	\\
%\hline    
%Czekanowski & 0	& 1	& 0	& 1	& 0	& 1	& 0	& 0	& 0	& 0	& 0	& 1	& 0	& 0	& 0	& 0	& 0	& 0	& 1	\\
%\hline    
%Kulczynski	& 0	& 1	& 0	& 1	& 0	& 1	& 0	& 0	& 0	& 0	& 0	& 0	& 0	& 0	& 0	& 0	& 0	& 0	& 1	\\
%\hline
%sub-set	   & 0? & 1 & 0 & 1? & 0 & 1? & 0	& 0	& 0	& 0	& 0	& ?	& 0	& 0	& 0	& 0? & 0 & 0 & 1 \\
%\hline
%\end{tabular}
%\end{table}

\begin{table}
\centering
\begin{tabular}{|c|c|c|c|c|c|c|c|}
%\scriptsize
\hline 
\textcolor{blue}{Prop} & \textcolor{blue}{Jac} & \textcolor{blue}{Supp} & \textcolor{blue}{Cos} & \textcolor{blue}{Rec} & \textcolor{blue}{Czek} & \textcolor{blue}{Kulc} & \textcolor{blue}{Sub-set} \\ 
\hline 
$P_{1}$ & \textcolor{blue}{0} & \textcolor{blue}{0} & \textcolor{blue}{0} & \textcolor{magenta}{1} & \textcolor{blue}{0} & \textcolor{blue}{0} & \textcolor{blue}{0?} \\ 
\hline 
$P_{2}$ & \textcolor{magenta}{1} & \textcolor{magenta}{1} & \textcolor{magenta}{1} & \textcolor{magenta}{1} & \textcolor{magenta}{1} & \textcolor{magenta}{1} & \textcolor{magenta}{1} \\ 
\hline 
$P_{3}$ & \textcolor{blue}{0} & \textcolor{blue}{0} & \textcolor{blue}{0} & \textcolor{blue}{0} & \textcolor{blue}{0} & \textcolor{blue}{0} & \textcolor{blue}{0} \\ 
\hline 
$P_{4}$ & \textcolor{magenta}{1} & \textcolor{blue}{0} & \textcolor{magenta}{1} & \textcolor{magenta}{1} & \textcolor{magenta}{1} & \textcolor{magenta}{1} & \textcolor{magenta}{1?} \\ 
\hline 
$P_{5}$ & \textcolor{blue}{0} & \textcolor{blue}{0} & \textcolor{blue}{0} & \textcolor{blue}{0} & \textcolor{blue}{0} & \textcolor{blue}{0} & \textcolor{blue}{0} \\ 
\hline 
$P_{6}$ & \textcolor{magenta}{1} & \textcolor{blue}{0} & \textcolor{magenta}{1} & \textcolor{magenta}{1} & \textcolor{magenta}{1} & \textcolor{magenta}{1} & \textcolor{magenta}{1?} \\ 
\hline 
$P_{7}$ & \textcolor{blue}{0} & \textcolor{blue}{0} & \textcolor{blue}{0} & \textcolor{blue}{0} & \textcolor{blue}{0} & \textcolor{blue}{0} & \textcolor{blue}{0} \\ 
\hline 
$P_{8}$ & \textcolor{blue}{0} & \textcolor{blue}{0} & \textcolor{blue}{0} & \textcolor{blue}{0} & \textcolor{blue}{0} & \textcolor{blue}{0} & \textcolor{blue}{0} \\ 
\hline 
$P_{9}$ & \textcolor{blue}{0} & \textcolor{blue}{0} & \textcolor{blue}{0} & \textcolor{blue}{0} & \textcolor{blue}{0} & \textcolor{blue}{0} & \textcolor{blue}{0} \\ 
\hline 
$P_{10}$ & \textcolor{blue}{0} & \textcolor{blue}{0} & \textcolor{blue}{0} & \textcolor{blue}{0} & \textcolor{blue}{0} & \textcolor{blue}{0} & \textcolor{blue}{0} \\ 
\hline 
$P_{11}$ & \textcolor{blue}{0} & \textcolor{blue}{0} & \textcolor{blue}{0} & \textcolor{blue}{0} & \textcolor{blue}{0} & \textcolor{blue}{0} & \textcolor{blue}{0} \\ 
\hline 
$P_{12}$ & \textcolor{blue}{0} & \textcolor{magenta}{1} & \textcolor{magenta}{1} & \textcolor{magenta}{1} & \textcolor{magenta}{1} & 0 & ? \\ 
\hline 
$P_{13}$ & \textcolor{blue}{0} & \textcolor{blue}{0} & \textcolor{blue}{0} & \textcolor{blue}{0} & \textcolor{blue}{0} & \textcolor{blue}{0} & \textcolor{blue}{0} \\ 
\hline 
$P_{14}$ & \textcolor{blue}{0} & \textcolor{blue}{0} & \textcolor{blue}{0} & \textcolor{blue}{0} & \textcolor{blue}{0} & \textcolor{blue}{0} & \textcolor{blue}{0} \\ 
\hline 
$P_{15}$ & \textcolor{blue}{0} & \textcolor{blue}{0} & \textcolor{blue}{0} & \textcolor{blue}{0} & \textcolor{blue}{0} & \textcolor{blue}{0} & \textcolor{blue}{0} \\ 
\hline 
$P_{16}$ & \textcolor{blue}{0} & \textcolor{blue}{0} & \textcolor{blue}{0} & \textcolor{blue}{0} & \textcolor{blue}{0} & \textcolor{blue}{0} & \textcolor{blue}{0?} \\ 
\hline 
$P_{17}$ & \textcolor{blue}{0} & \textcolor{blue}{0} & \textcolor{blue}{0} & \textcolor{blue}{0} & \textcolor{blue}{0}0 & \textcolor{blue}{0} & \textcolor{blue}{0} \\ 
\hline 
$P_{18}$ & \textcolor{blue}{0} & \textcolor{blue}{0} & \textcolor{blue}{0} & \textcolor{blue}{0} & \textcolor{blue}{0} & \textcolor{blue}{0} & \textcolor{blue}{0} \\ 
\hline 
$P_{19}$ & \textcolor{magenta}{1} & \textcolor{magenta}{1} & \textcolor{magenta}{1} & \textcolor{magenta}{1} & \textcolor{magenta}{1} & \textcolor{magenta}{1} & \textcolor{magenta}{1} \\ 
\hline 
\end{tabular} 
\caption{Evaluation of properties on a subset of measures of Class 4.}
\label{tab:evalc4}
\end{table}

\begin{figure}
\centering
\makebox[0cm][r]{\includegraphics[width=0.45\linewidth]{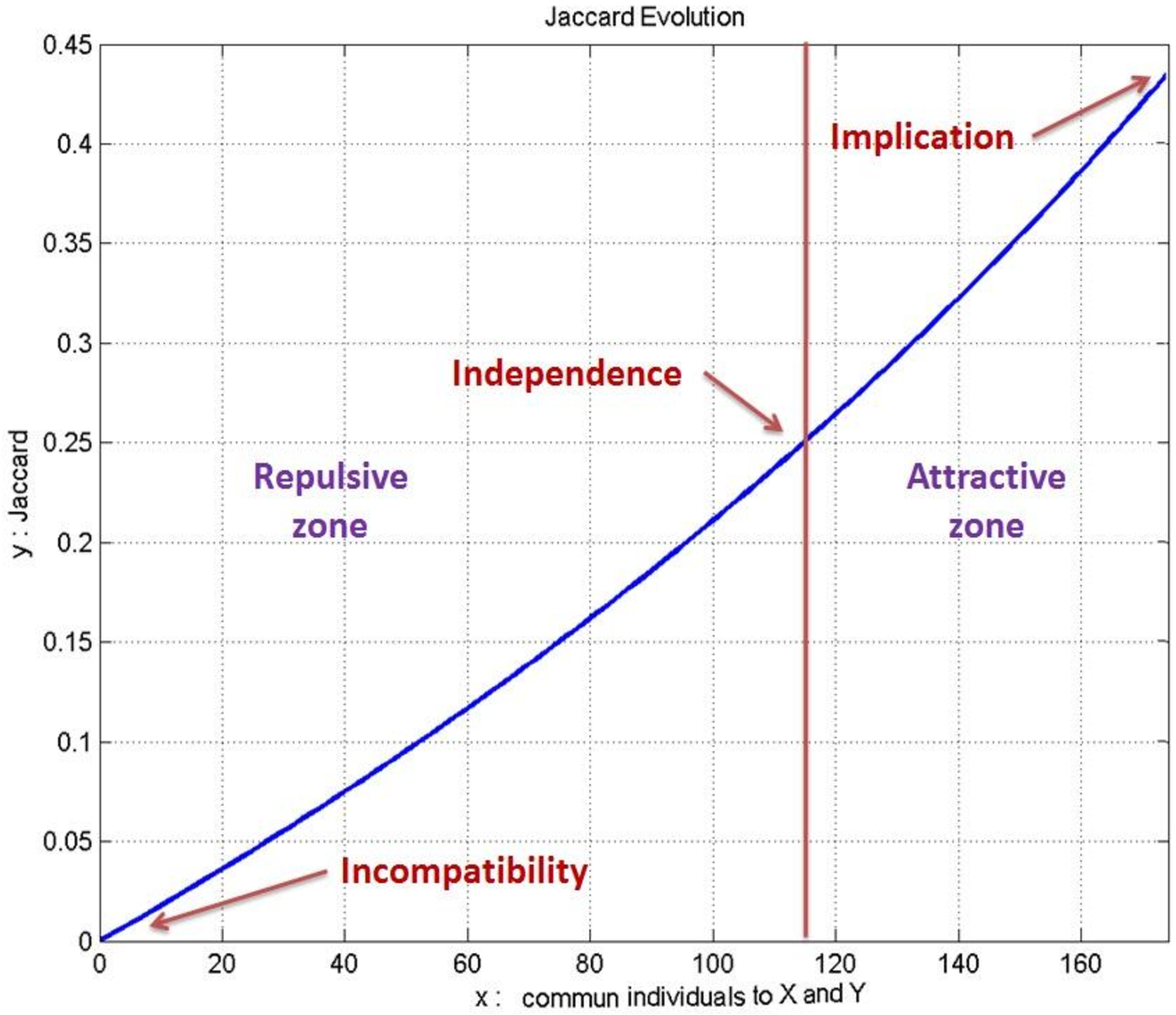}}
\hspace{0.7cm}
\makebox[0cm][l]{\vspace{0.5\linewidth}\includegraphics[width=0.45\linewidth]{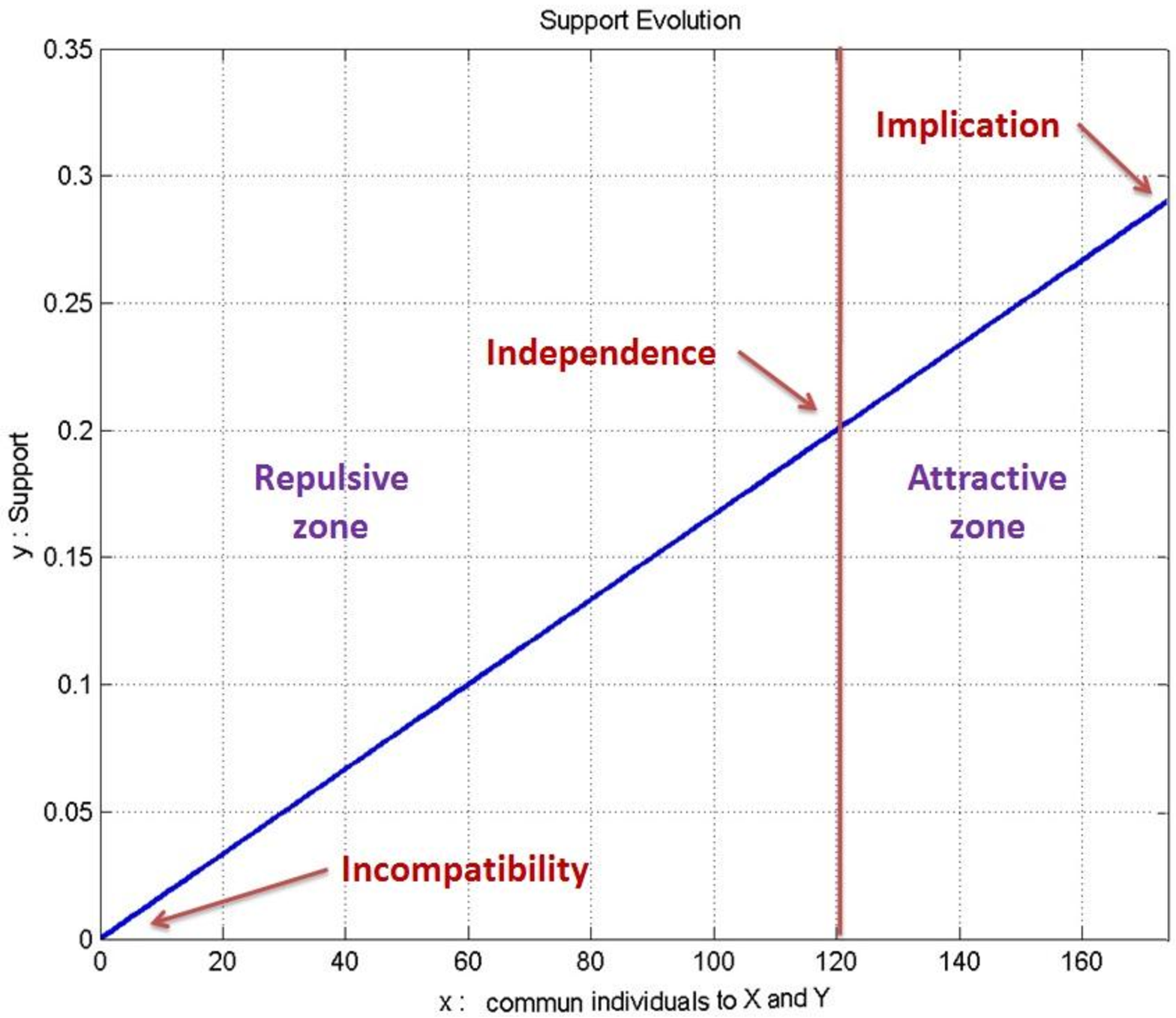}}\\
\vspace{0.35cm}
\makebox[0cm][r]{\includegraphics[width=0.42\linewidth]{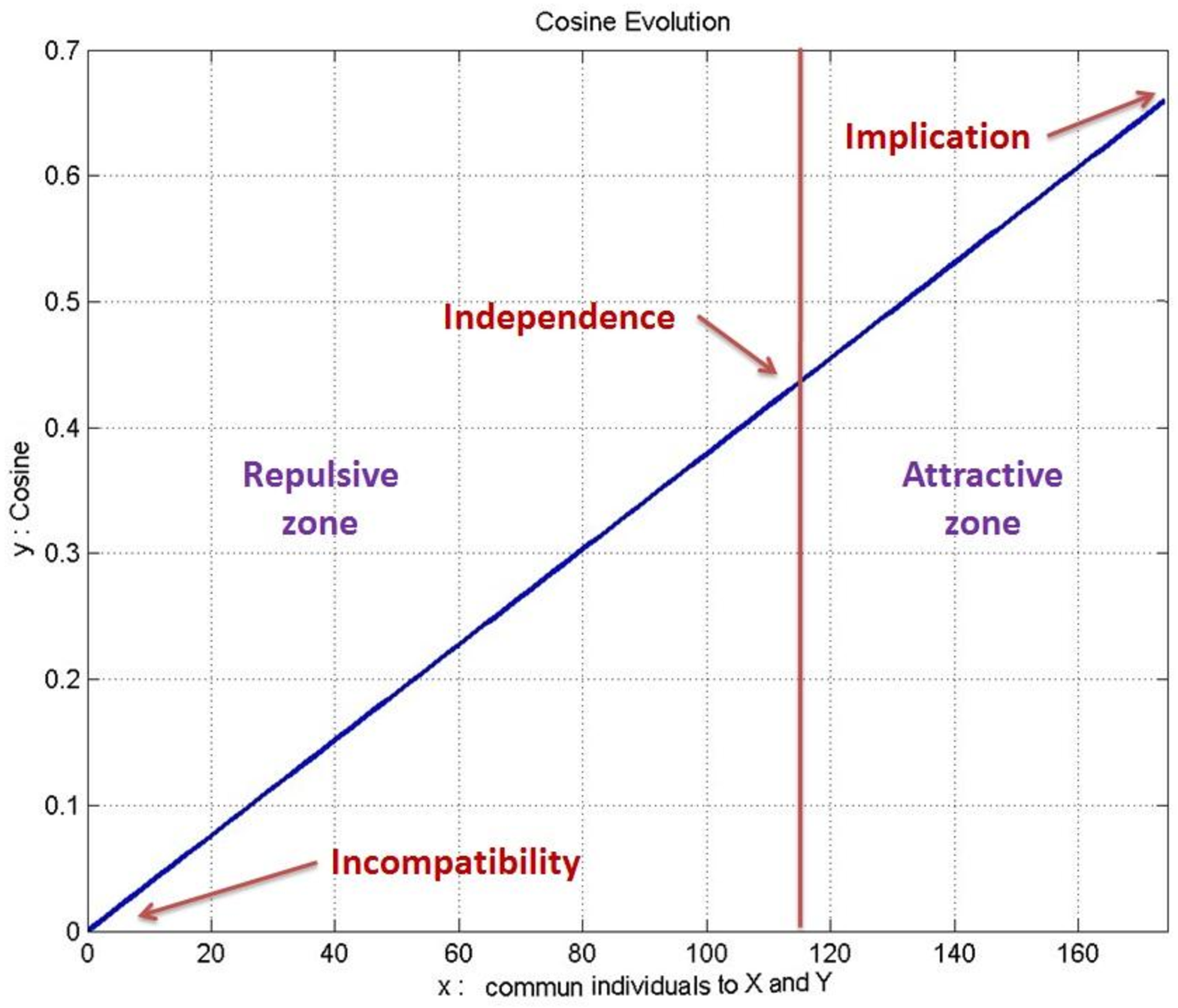}}
\hspace{0.7cm}
\makebox[0cm][l]{\vspace{0.4\linewidth}\includegraphics[width=0.47\linewidth]{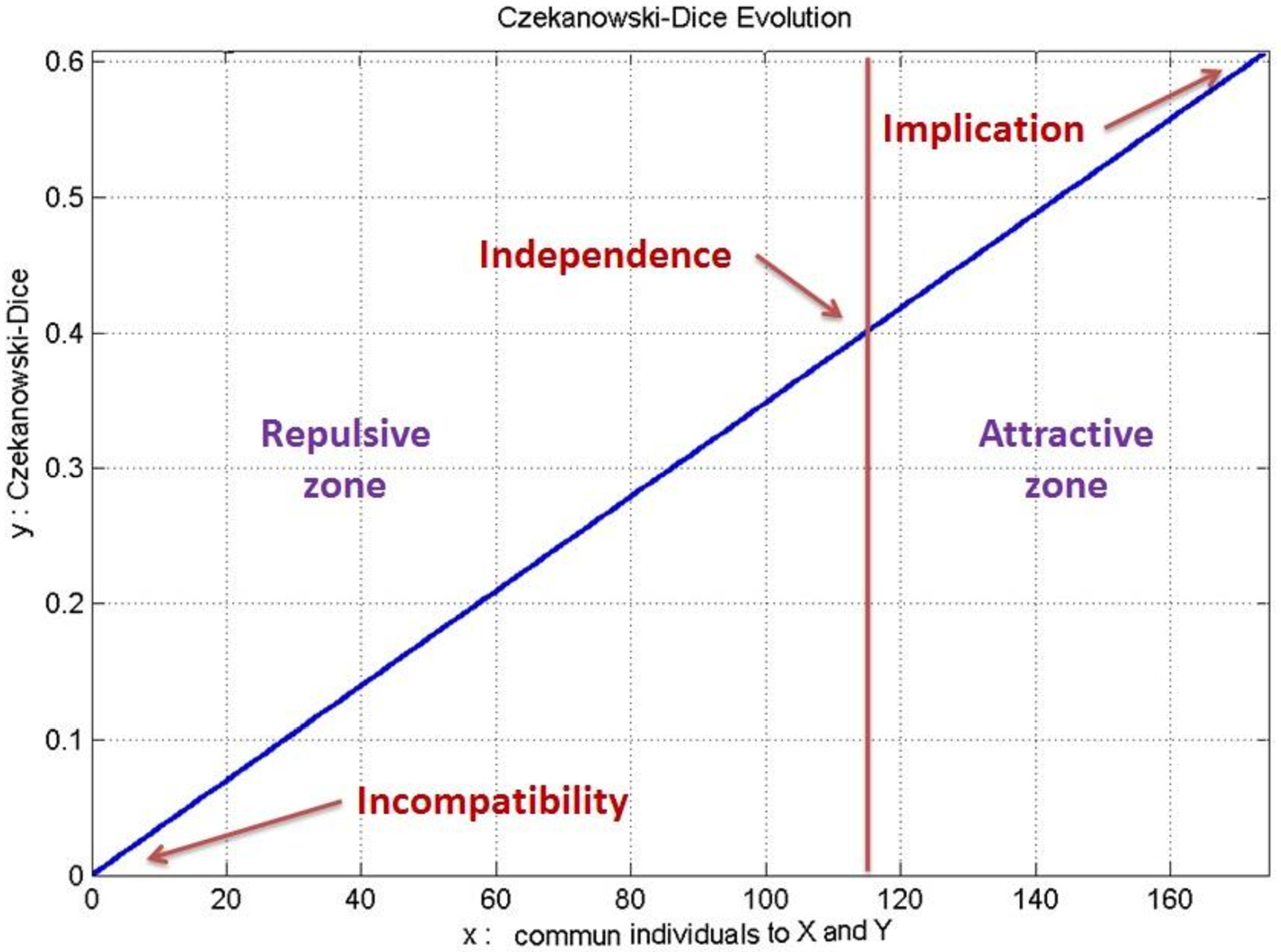}}\\
\vspace{0.35cm}
\makebox[0cm][r]{\includegraphics[width=0.44\linewidth]{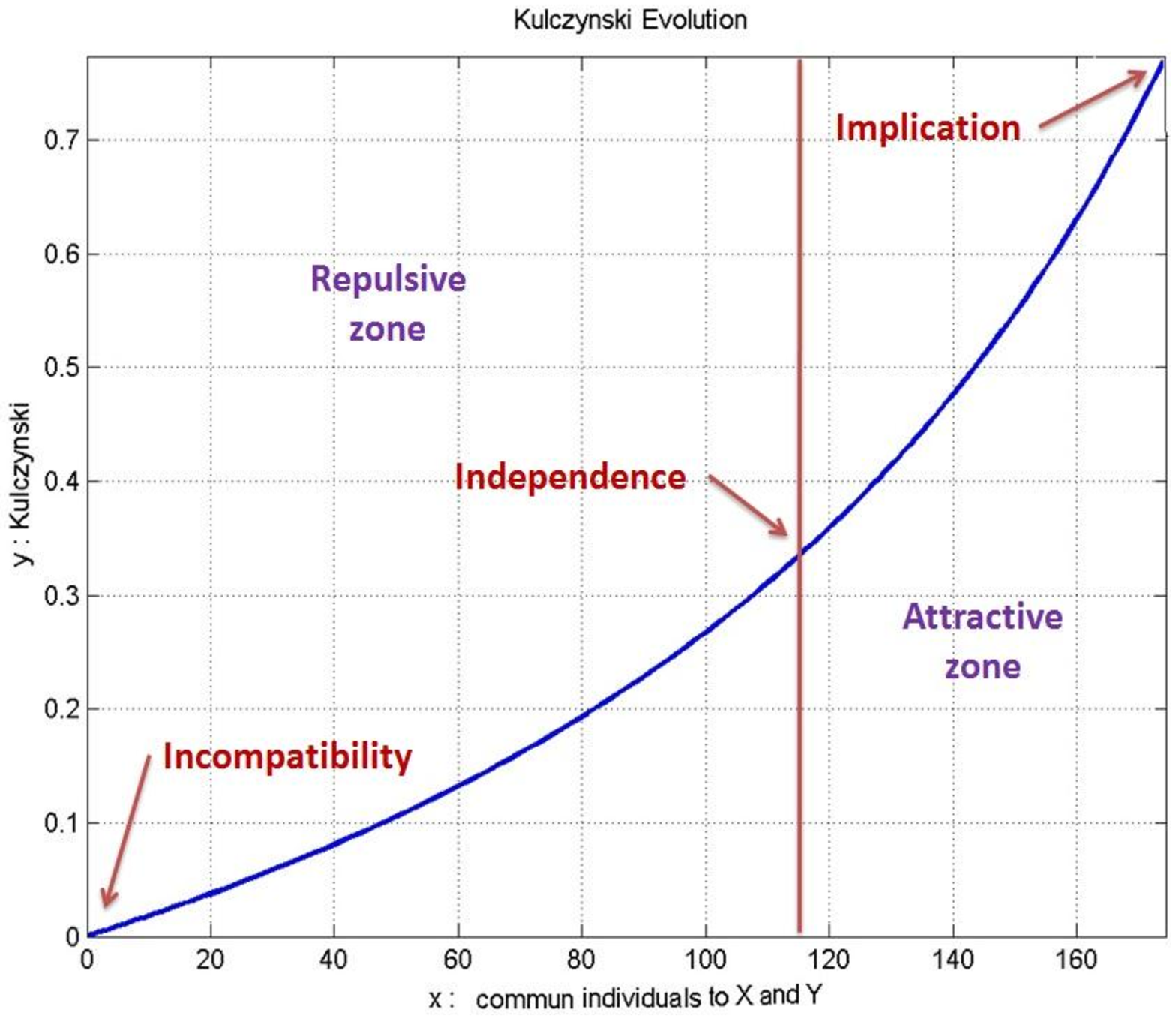}}
\hspace{0.7cm}
\makebox[0cm][l]{\vspace{0.5\linewidth}\includegraphics[width=0.45\linewidth]{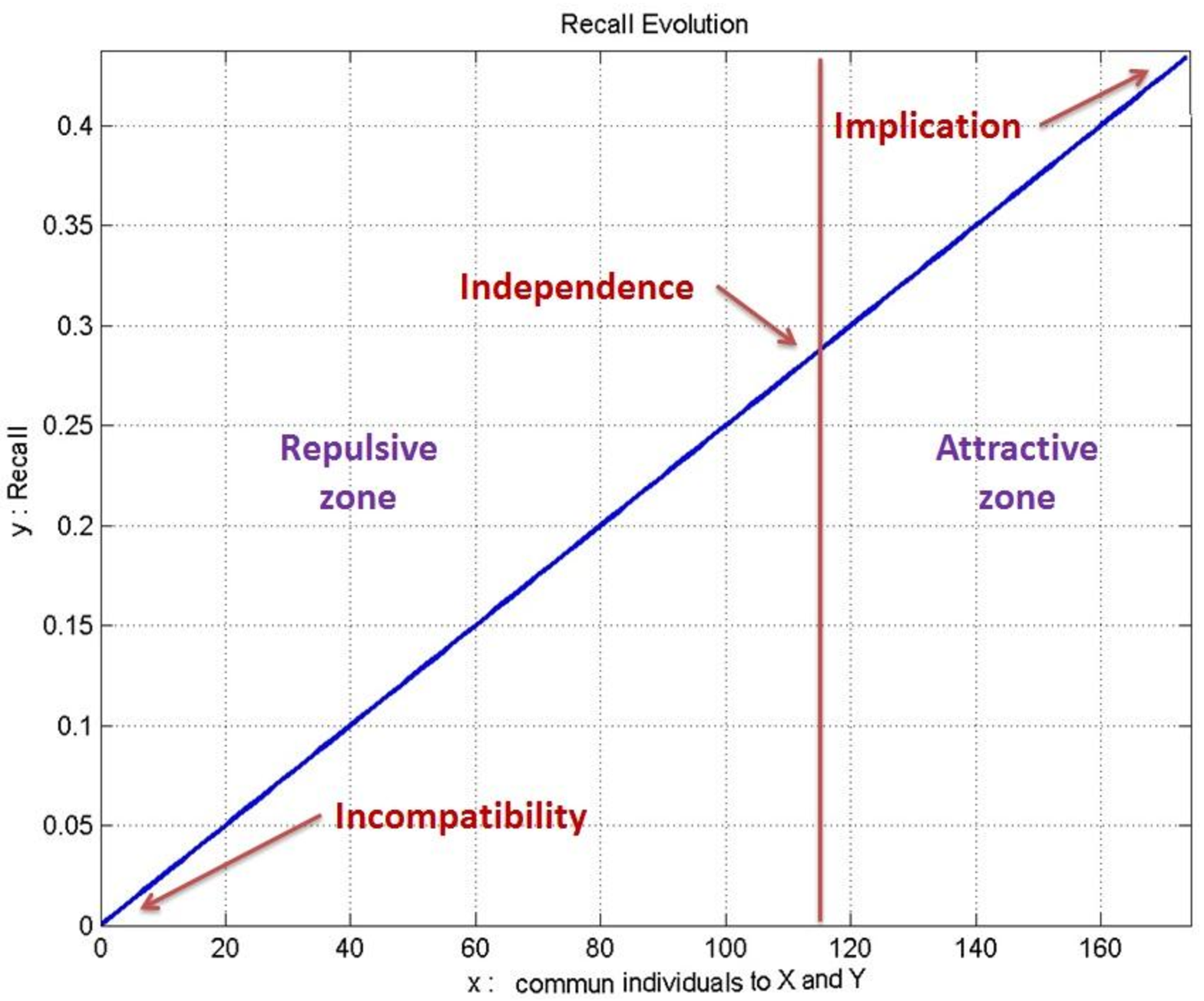}}
\vspace{0.3cm}
\caption{Evolution of the six measures of $C_{4}$ class according to the number of examples.}
\label{fig:C4Measures}
\end{figure}

As with the previous class $C_{6}$, we will study the evolution of these different measures according to the number of examples. {\em Figure\/}~\ref{fig:C4Measures} restitutes this evolution. We retained the same cardinality as above for the sets premise, conclusion and the whole data set ($n_{X} = 174$, $n_{Y} = 400$ and $n = 600$).\\

We test the null value taken by these measures in the case of incompatibility. We obtain two types of curves:

\begin{itemize}
\item A straight line for the measures {\em \textcolor[RGB]{127, 0, 255}{Support}, \textcolor[RGB]{127, 0, 255}{Cosine}, \textcolor[RGB]{127, 0, 255}{Czekanowski-Dice}\/} and {\em \textcolor[RGB]{127, 0, 255}{Recall}\/},
\item A half-parabole for the measures {\em \textcolor[RGB]{127, 0, 255}{Jaccard}\/} and {\em\textcolor[RGB]{127, 0, 255}{ Kulczynski}\/}.
\end{itemize}

After studying more precisely some classes and tried to give an interpretation to them, now we validate our work by a comparison with existing classifications \cite{Vaillant06}, \cite{Lesot:2010:OED}, \cite{ZighedAB11}, \cite{HeraviZ10}.

\section{Validation}
\label{sec:valid}

Many studies have been realized to better understand interestingness measure behavior. In this section, we want to compare clusters of measures we obtained with previous works \cite{Vaillant06}, \cite{HuynhGB07}, \cite{LeBras2011}, \cite{Lesot:2010:OED}, \cite{ZighedAB11} in order to highlight similarities and differences between them.

\subsection{Comparison with the work of B. Vaillant} \label{subsec: valid_Vaillant}

We first compare the classification we obtained with that of Beno\^it Vaillant \cite{Vaillant06}, who made his study on $20$ measures according to $9$ formal properties. From these nine properties, we have $7$ properties in common as "{\em comprehensability of the measure\/}" and "{\em easiness to set a threshold of acceptance\/}" are considered too subjective. To perform a classification, Beno\^it Vaillant also used Ward criterion but has retained Manhattan distance. The author points out that by using other criteria, he obtained similar results. He identified the five following classes: 

\begin{itemize}
\item \textcolor[RGB]{127, 0, 255}{{\em $ClBV_{1}$ = \{Support, Least contradiction, Laplace\}\/}},
\item \textcolor[RGB]{127, 0, 255}{{\em $ClBV_{2}$ = \{Confidence, Sebag, Examples rate\}\/}},
\item \textcolor[RGB]{127, 0, 255}{{\em $ClBV_{3}$ = \{Correlation coefficient, Piatetsky-Shapiro, Pavilion, Interest, Implication index, Cohen, Informational gain\}\/}},
\item \textcolor[RGB]{127, 0, 255}{{\em $ClBV_{4}$ = \{Loevinger, Bayes factor, Conviction\}\/}} and
\item \textcolor[RGB]{127, 0, 255}{{\em $ClBV_{5}$ = \{Zhang, IIET, Intensity of implication, Probabilistic Discriminant index\}\/}}.
\end{itemize}

We can assimilate {\em IIET\/} measure with {\em REII\/} because the purpose of both of them is the same. 

We agree on the following grouping: 

\textcolor[RGB]{127, 0, 255}{$ClBV_{2} \subset C_{5}$, $ClBV_{4} \subset C_{7}$},

and we have the following relations between groups:
\textcolor[RGB]{127, 0, 255}{
$ClBV_{1} - {Support} \subset C_{5}$, $ClBV_{3} - {Implication~index} \subset Gp_{8}~\cup ~C_{7}$ and
$ClBV_{5} - {Zhang} \subset C_{1} \cup C_{2}$}.

The grouping where the disagreement is the most important is \textcolor[RGB]{127, 0, 255}{$ClBV_{3}$}, since we have bring up \textcolor[RGB]{127, 0, 255}{$ Gp_{8} $} group which is present with only one technique: a version of {\em k\/}-means. As for \textcolor[RGB]{127, 0, 255}{$ClBV_{5}$} group, it includes all intensity of implication family measures, except {\em \textcolor[RGB]{127, 0, 255}{Zhang}\/} measure.

We studied $12$ additional properties, which explains why we do not find all the results of Beno\^it Vaillant.

%%Dois-je la laisser parce qu'ils réalisent une étude sur des datasets, ce n'est pas formel!!!!

In the following, we compare our results with those obtained by Y. Le Bras \cite{LeBras2011}.

\subsection{Comparison with the work of Y. Le Bras}\label{subsec:valid_LeBras}

In his work, Y. Le Bras \cite{LeBras2011} seeks to find common characteristics of objective measures. For that, he studied 42 interestingness measures according to six operational criteria that he proposed. These criteria concern from one hand the possibility to calculate robustness, and secondly to use efficient algorithms. Criteria are listed below:

\begin{itemize}
\item \textbf{Robustness measure calculation:} This is a rule measure resisting test w.r.t database disturbance \cite{BrasMLL10:rmar}
  \begin{enumerate}
    \item \textit{Planar measure}: for some measures, distance calculation is reduced to the calculation of the distance to a plan, which allows to provide exact algebric solution;
    \item \textit{Quadratic measure}: measures require to use a certain number of mathematical tools.
  \end{enumerate}
\item \textbf{Algorithmic properties} allowing algorithm to be effective:
  \begin{enumerate}
    \setcounter{enumi}2
    \item \textit{GUEUC}: it is the general property UEUC (\textit{Universal Existential Upward Closure}), which is a down monotonicity property;
    \item Measure \textit{omni-monotony}; 
    \item Measure \textit{opti-monotony};
  \end{enumerate}
\item \textbf{Anti-monotony property} of a measure for finding optimal rules:
  \begin{enumerate}
    \setcounter{enumi}5
    \item Measure \textit{anti-monotony}.
  \end{enumerate}
%\end{enumerate}
\end{itemize}

For each of the algorithmic properties, a generalization has been provided by the author (\textit{GUEUC}, \textit{omni-monotony} and \textit{opti-monotony}) as he proposed existence conditions of these generalizations .

By looking to the 6 described properties, we find that we are in total disagreement with Y. Le Bras w.r.t criteria choosen for studying measures behavior. Nevertheless, this does not prevent us from confronting our two works for a better understanding of measures behavior. In total, we have 38 common measures, some of them have the same definition but with different names \footnote{Interest represents Pearl in our work, Levier represents Novelty measure and J1-measure is Two-way support measure}. By comparing our works, we seek to identify whether common measures which belong to the same group evaluate properties studied by \cite{LeBras2011} in the same way.

The comparison of our results (\textit{section \ref{sec:fc}}) with those obtained by Y. Le Bras reveals similarities according to these groups of measures.

\begin{itemize}
\item \textcolor{blue}{$C_{3}$}: clusters Coverage, Gini, Implication index, J-measure, Prevalence and Pearl (which according to k-means, belongs to this group) measures, common to both works. According to \cite{LeBras2011}, none of these measures is quadratic or anti-monotonic. He also shows the closeness of Coverage and Prevalence measures, since they are the only two planar and omni-monotonic measures having GUEUC property;

\item \textcolor{blue}{$C_{4}$}: contains the following common measures Cosine, Czekanowski-dice, Jaccard, Kulczynski, Accuracy, Specificity, Support and Recall. All of them, except Cosine which is quadratic, are planar and verify the antimonotony property. Furthermore, we find that most of these measures have GUEUC property, except Kulczynski and Specificity. Support is the only omni-monotonic measure in this cluster;

\item \textcolor{blue}{$C_{5}$}: we remark that Descriptive confirmation is the only measure absent from this group. Y. Le Bras's work \cite{LeBras2011} reveals that Examples rate, Sebag, Ganascia and Confidence verify all the studied properties in the same way. Otherwise, none of $C_{5}$ commun measures is quadratic, but they all are omni and opti-monotonic. We realize also that all these measures are planar, except Laplace and that only two of them (Least contradiction and Laplace) are anti-monotonic;

\item \textcolor{blue}{$C_{6}$}: we find the following three opti-monotonic measures Y, Q Yule and Zhang, which do not check any of these properties, antimonotony, omni-monotony and planar measure. Visualizing the behavior of Piatetsky-shapiro and Novelty measures, which belong to this group according to the hierarchical method, we find that they are also opti-monotone and do not check the omni-monotony property and planar measure. Novelty, which seems to be more robust than Piatetsky-Shapiro (it is quadratic), is the only measure which has the good property of anti-monotonicity in class rules case;

\item \textcolor{blue}{$C_{7}$}: all $C_{7}$ measures have been studied by \cite{LeBras2011}, including Collective strength, Cohen and Odds ratio measures, which according to the hierarchical method belong to $ C_{7} $. Among all these measures, only Cohen is anti-monotonic, but none of them is omni-monotonic or planar. GUEUC property is verified by Pavilion, Conviction, Factor of Bayes, Informational gain, Interest and Loevinger, which are quadratic and opti-monotonic, identifying then strong operational properties with Cohen, Odds ratio and Relative risk measures.
\end{itemize}

Following our works comparison, we notice that from Y. Le Bras study on interestingness measures according to the six proposed criteria, we can identify behavior similarities between common measures of the same group. The only group which doesn't reveal a good agreement is \textcolor{blue}{$ C_{3}$}.

Another classification realized by \cite{HuynhGB07} on interestingness measures using datasets is presented in the next section and compared with the classification obtained in \textit{section \ref{sec:fc}}.

\subsection{Comparison with the work of Hyunh et al.} \label{subsec: valid_Hyunh}

Another classification was made by Huynh et al. \cite{HuynhGB07}, who studied $36$ interestingness measures, with $32$ commun measures, on $2$ datasets with opposite nature: one highly correlated (mushroom) and the other weakly correlated synthetic base (T5.I2.D10K). authors present initially a taxonomy of measures according to the following $2$ criteria: \ \

\begin{enumerate}
   \item \textit{Topic}: deviation from independence or equilibrium;
   \item \textit{Nature}: descriptive or statistical.
\end{enumerate}

From the study of these two particular parameters on datasets, the $5$ following groups of measures are retained:

\begin{itemize}
    \item \textcolor{blue}{$Cl_{de}$} (\textit{descriptive / deviation from equilibrium}): \{\textit{Confidence, Laplace, Sebag, Examples rate, Descriptive confirmation, Descriptive confirmed-confidence, Least contradiction} \};
    \item \textcolor{blue}{$Cl_{di}$} (\textit{descriptive / deviation from independence}): \{\textit{Correlation, Interest, Loevinger, Conviction, Dependency, Pavillon, J-measure, Gini, TIC, Collective strength, Odds ratio, Yule's Q, Yule's Y, Klosgen, Cohen} \};
    \item \textcolor{blue}{$Cl_{se}$} (\textit{Statistical / deviation from equilibrium}): \{\textit{IPEE \}};
    \item \textcolor{blue}{$Cl_{si}$} (\textit{Statistical / deviation from independence}): \{\textit{II, EII, EII2, Lerman, Interest Rule} \};
    \item \textcolor{blue}{$Cl_{o}$} (\textit{other}): \{\textit{Support, Precision, Jaccard, Cosine, Causal confidence, Causal confirmation, Causal confirmed-confidence, Causal dependency} \}. \\
\end{itemize}

By comparing these $5$ groups of measures with those described in Figure \ref{fig:ClassificationFormelle}, we note our agreement on the categorization of the following measures: \textcolor[RGB]{127, 0, 255}{\{\textit{Confidence, Laplace, Sebag, Examples rate, Least contradiction} \} $\subset C_{5}$}, \textcolor[RGB]{127, 0, 255}{\{\textit{Correlation, Cohen, Collective strength, Odds ratio} \} $\subset Gp_{8}$} since they are gathered according to the partitioning method \textit{K-means}, \textcolor[RGB]{127, 0, 255}{\{\textit{Gini, J-measure, Dependence, Klosgen} \} $\subset C_{3}$}, \textcolor[RGB]{127, 0, 255}{\{\textit{Interest, Loevinger, Conviction, Pavilion, Klosgen \}} $\subset C_{7}$}, \textcolor[RGB]{127, 0, 255}{\{\textit{Yule's Q, Yule's Y} \} $\subset C_{6}$} and finally \textcolor[RGB]{127, 0, 255}{\{\textit{Jaccard, Cosine, Causal confirmation, Causal Confidence, Causal confirmed-confidence} \} $\subset C_{4}$}. According to this comparison, we highlight similarities between groups of common measures revealed by both works.

\subsection{Comparison with other works} \label{subsec: valid_others}

Another classification was performed on distance and similarity measures by Marie-Jeanne Lesot and Maria Rifgi \cite{Lesot:2010:OED}. Authors studied the induced order using measures and not the obtained numerical values, since their context of study is the information research. This study focused on measures dedicated to binary and digital data by conducting experiments on both real and artificial data. The authors obtained a list of equivalent measures (measures that induce always the same order) and for non-equivalent measures, they quantified the disagreement by a degree of equivalence based on the generalized Kendall's coefficient. On the $10$ measures studied and designed for binary data, five are common to our two studies. These measures are: {\em \textcolor[RGB]{127, 0, 255}{Czekanowski-Dice}, \textcolor[RGB]{127, 0, 255}{Jaccard}, \textcolor[RGB]{127, 0, 255}{Ochiai}, \textcolor[RGB]{127, 0, 255}{Yule's Y}\/} and {\em \textcolor[RGB]{127, 0, 255}{Yule's Q}\/}. Authors found that {\em Yule's Y\/} and {\em Yule's Q\/} are equivalent measures. This result is also confirmed by our study since these two measures are in the same class $ C_{6} $ as we have already mentioned, and are very close according to the dendogram of the {\em figure\/}~\ref{fig:dendogram}. They also found that {\em \textcolor[RGB]{127, 0, 255}{Czekanowski-Dice}\/} and {\em \textcolor[RGB]{127, 0, 255}{Jaccard}\/} are equivalent measures. Both measures were also assigned to the same class: the class $ C_{4} $, and we find them with a relatively large proximity in the dendogram of the {\em figure\/}~\ref{fig:dendogram} (we chose Cosine measure as a representative one on the dendrogram as we have discussed in {\em Section\/}~\ref{sec:evalprop}, {\em Cosine\/} and {\em Czekanowski-Dice\/} measures have identical values for the $19$ properties which led to the formation of $ G_{3} $ group). Finally, we grouped also {\em \textcolor[RGB]{127, 0, 255}{Ochiai}\/} (or {\em \textcolor[RGB]{127, 0, 255}{Cosine}\/}) measures with {\em \textcolor[RGB]{127, 0, 255}{Czekanowski-Dice}\/} and {\em \textcolor[RGB]{127, 0, 255}{Jaccard}\/} in $ C_{4} $ cluster. Authors \cite{Lesot:2010:OED} found a degree of equivalence between Ochiai measure and the equivalence class {{\em \textcolor[RGB]{127, 0, 255}{Czekanowski-Dice}, \textcolor[RGB]{127, 0, 255}{Jaccard}\/}} of $0.99$, which confirms our results. 

A final classification was proposed by Djamel Zighed, Rafik Abdesselam and Ahmed Bounekkar \cite{ZighedAB11} on $13$ proximity measures. Only two of them are common to our two studies: {\em \textcolor[RGB]{127, 0, 255}{Cosine}\/} and {\em \textcolor[RGB]{127, 0, 255}{Correlation coefficient}\/}. The classification they proposed is based on the topological equivalence and uses the structure of local neighborhood. Both measures appeared very close in this classification, in contrast to our work as we find them in classes $ C_{4} $ and $ Gp_{8} $. The set of studied measures are so different, the founded classes by each technique are difficult to compare. Moreover, as authors emphasized during the presentation of their work, the classification they obtained is performed as poorly representative because it is applied on a single dataset: Fisher's Iris.

We are well aware that measures categorization may also depend on several factors including: the data, the expert user, the nature of the extracted rules and classes search procedure, as highlighted by \cite{Suzuki08}.
To avoid bias data, the expert and the nature of the extracted rules, we have chosen here a theoretical study based on properties of measures, rather than experimental data \cite{HuynhGB05:cimpc}. Both are obviously complementary.

To avoid the bias of the clusters construction procedure, we used two classification techniques, which generally exhibited strong similarities between many measures, and highlights similarities and differences with previous works (\cite{Vaillant06}, \cite{Lesot:2010:OED}, \cite{ZighedAB11}.
This study complements previous works on the description of a unifying vision of interestingness measures \cite{Hebert2007}, and adds a further contribution to the analysis of these measures.

\section{Conclusion}
\label{sec:con}

This article takes as its starting point a synthesis paper on interestingness measures present in the literature to extract knowledge and properties judged relevant to them. This synthesis work led to the assessment of $19$ properties judged interesting on $61$ measures. The objective of this paper is the classification of these measures to assist the user in his choice of complementary measures to the couple ({\em Support, Confidence\/}) to eliminate uninteresting rules. Initially, we analyzed these data (matrix of $61~measures \times 19~properties$) to determine if simplification was not feasible by looking first to groups of measures with completely identical behavior and then by detecting if properties were not redundant. We detected seven groups of measures with completely identical behavior which enabled to reduce our starting data for the classification research by two techniques: a method of agglomerative hierarchical classification and a version of {\em k \/}-means method. Classifications obtained from both techniques allowed to reach a consensus: $7$ classes were partially validated by existing classifications. 

In the future, we would like to consolidate classes of measures we obtained by comparing the {\em N\/} best extracted rules in different databases and by each of the studied measures to verify that this set of {\em N\/} best rules is substantially the same in each class. Finally, it would be interesting to consider smaller classes ({\em with the help of the extracted dendogram\/}) to assign a semantic to each of them, which would be a great help to the user ({\em rather than a set of verified properties\/}), since we saw our inability to define in a few words or phrases each of these extracted classes. Complementary properties might to be considered. The notion of association rules robustness \cite{BrasMLL10:rmar} could be also considered in the interestingness measures categorization.

\begin{acknowledgements}
We thank Israël-César Lerman for his constructive comments on this article. Moreover, this work is partially supported by the French-Tunisian PHC Utique 11G1417: EXQUI ({\em EXtraction, QUality and Knowledge Engineering in heterogeneous environments\/}). 
\end{acknowledgements}

\section{Annexe 1}
\label{sec:annexe1}

%\begin{table}
\begin{center}
\scriptsize
\begin{longtable}{|p{0,3cm}|p{2cm}|p{7,72cm}|}
%\begin{tabular}{|p{0,4cm}|p{2cm}|p{5cm}|}
\hline 
\textcolor{blue}{\textbf{$N^{\circ}$}} & \textcolor{blue}{\textbf{Measure}} & \textcolor{blue}{\textbf{Formula}} \\ 
\hline 
1 & Correlation coefficient & $\frac {p(XY) - p(X) p(Y)} { \sqrt {p(X) p(Y)p(\bar{X}) p(\bar{Y})}  } $ \\ 
\hline 
2 & Cohen or Kappa & $  2~ \frac { p(XY) - p(X) p(Y)}{p(X) + p(Y) - 2  p(X) p(Y)} $ \\ 
\hline 
3 &  Confidence or precision & $\frac{p(XY)}{p(X)} $ \\ 
\hline 
4 &  Causal Confidence & $1 - \frac{1}{2}\left(\frac{1}{p(X)} + \frac{1}{p(\bar{Y})}\right)p(X\bar{Y})$ \\ 
\hline 
5 &  Centered Confidence or Pavillon & $ \frac{p(XY)}{p(Y)} - p(Y)$ \\ 
\hline 
6 &  Descriptive Confirm Confidence or Ganascia & $ 1 - 2~\frac{p(X\bar{Y})}{p(X)} $ \\ 
\hline 
7 & Causal Confirm Confidence & $1 - \frac{1}{2}\left(\frac{3}{p(X)} + \frac{1}{p(\bar{Y})}\right)p(X\bar{Y}) $ \\ 
\hline 
8 & Causal Confirm & $p(X) + p(\bar{Y}) - 4p(X\bar{Y})$ \\ 
\hline 
9 & Descriptive Confirm & $p(XY)  - p(X\bar{Y})  $ \\ 
\hline 
10 &  Conviction & $\frac{p(X)p(\bar{Y})}{p(X\bar{Y})} $ \\ 
\hline 
11 &  Cosinus or Ochiai & $\frac{p(XY)}{\sqrt{p(X)p(Y)}} $ \\ 
\hline 
12 &  Coverage & $p(X) $ \\ 
\hline 
13 & Czekanowski-Dice or F-measure & $2~\frac{p(XY)}{p(XY) + 1 - p(\bar{X}\bar{Y})} $ \\ 
\hline 
14 & Dependency & $\| p(\bar{Y}) - \frac{p(X\bar{Y})}{p(X)}\|$ \\ 
\hline 
15 & Putative Causal Dependency & $ \frac{3} {2}  + 2p(X) - \frac{3} {2}p(Y) - \Big( \frac{3} {2p(X)} +
                                                       \frac{2} {p(\bar{Y})} \Big) p(X\bar{Y})$ \\ 
\hline 
16 & Gray and Orlowska's Interestingness Weighting Dependency & $ \Bigg ( \Big (   \frac{p(XY)}{p(X)p(Y)}  \Big )^k - 1    \Bigg )  \times p(XY)^m$ \\ 
\hline 
17 & Bayes factor or Odd multiplier & $\frac{p(XY) p(\bar{Y})}{p(X\bar{Y})p(Y)} $ \\ 
\hline 
18 & Certainty factor or Loevinger or Satisfaction & $ \frac{p(XY) - p(X)p(Y)}{p(X)p(\bar{Y})} $  \\ 
\hline 
19 & Negative reliability & $\frac{p(\bar{X}\bar{Y})}{p(\bar{Y})} $ \\ 
\hline 
20 & Collective Strength & $\frac{p(XY) + \frac{p(\bar{X}\bar{Y})}{p(\bar{X})}}{p(X)p(Y) + p(\bar{X})p(\bar{Y})} \times
  \frac{1 - p(X)p(Y) - p(\bar{X})p(\bar{Y})}{1 - p(XY) - \frac{p(\bar{X}\bar{Y})}{p(\bar{X})}}$ \\ 
\hline 
21 &  Fukuda & $ n~ \Big ( p(XY) - \sigma_c~ p(X) \Big )$ \\ 
\hline 
22 & Informational gain & $ log_{2} \Big ( \frac {p(XY)} {p(X)p(Y)} \Big ) $ \\ 
\hline 
23 & Gini & $p(X)~\Big (  \frac{p^2(XY)}{p^2(X)} +  \frac{p^2(X\bar{Y})}{p^2(X)}  \Big )
        +  p(\bar{X}) ~\Big (  \frac{p^2(\bar{X}Y)}{p^2(\bar{X})} +   \frac{p^2(\bar{X}\bar{Y})}{p^2(\bar{X})}  \Big )
         -   p^2(Y) -   p^2(\bar{Y}) $ \\ 
\hline 
24 & Goodman-Kruskal & $ \frac{\sum_{j} max_{k} P(X_{j}, Y_{k}) + \sum_{k} max_{j} P(X_{j}, Y_{k}) - max_{j} P(X_{j}) - max_{k} P(Y_{k})}{2 - max_{j} P(X_{j}) - max_{k} P(Y_{k}) }  \times 
\frac{\frac{n_{XY}n_{\bar{XY}}}{n^{2}} -  \frac{n_{X\bar{Y}}n_{\bar{X}Y}}{n^{2}}}{\frac{n_{XY}n_{\bar{XY}}}{n^{2}} + \frac{n_{X\bar{Y}}n_{\bar{X}Y}}{n^{2}}}$ \\ 
\hline 
25 & Implication index & $\sqrt{n} ~ \frac { p(X\bar{Y}) - p(X) p(\bar{Y})}{\sqrt{p(X) p(\bar{Y}) } } $ \\ 
\hline 
26 & Probabilistic intensity of deviation from equilibrium (IPEE) & $ P \Big [ N(0,1) \geq \frac{n_{X\bar{Y}} - n_{XY}}{\sqrt{n_{X}}} \Big ]$ \\ 
\hline 
27 & Entropic probabilistic intensity of deviation from equilibrium (IP3E) & $ \sqrt{\Big [\frac{1}{2} \Big (  (1- h_{1}(P(X\bar{Y}))^{2})  \times  (1- h_{2}(P(X\bar{Y}))^{2})    \Big)^\frac{1}{4}  + 1 \Big ]} \times \sqrt{IPEE} ~~
with~h_{1}(t) = - \Big ( 1 - \frac{t}{P(X)} \Big ) log_{2} \Big ( 1 - \frac{t}{P(X)} \Big ) - \frac{t}{P(X)} log_{2} \Big ( \frac{t}{P(X)} \Big )~for~t \in \Big [0, \frac{P(X)}{2}  \Big [,
~~else~~h_{1}(t) = 1~~
h_{2}(t) = - \Big ( 1 - \frac{t}{P(\bar{Y})} \Big ) log_{2} \Big ( 1 - \frac{t}{P(\bar{Y})} \Big ) - \frac{t}{P(\bar{Y})} log_{2} \Big ( \frac{t}{P(\bar{Y})} \Big )~for~t \in \Big [0, \frac{P(\bar{Y})}{2}  \Big [,
~~else~~h_{2}(t) = 1 $ \\

\hline 
28 & Probabilistic discriminant index (PDI) & $ P \Big [ N(0,1) \geq II^{CR/B} \Big ]$ where $ II^{CR/B} $ indicate that \textit{II} is reduced-centred according to the values taken by \textit{II} on the extracted rules set.\\ 
\hline 
29 & Mutual Information &  $\frac{VS(XY)}{-P(X)log_{2}P(X) - P(\bar{X})log_{2}P(\bar{X})}$\\ 
\hline 
30 & Intensity of Implication (II) & $  P \Big [ Poisson (n P(X) P(\bar{Y})) \geq  P(X\bar{Y})  \Big ]$ \\ 
\hline 
31 & Entropic intensity of implication (IIE) & $ \sqrt{ \Big [ \big ( 1 - h_{1}(P(X\bar{Y}))^{2} \big )  \times \big ( 1 - h_{2}(P(X\bar{Y}))^{2}   \big )  \Big ]^\frac{1}{4} \times II}$ \\ 
\hline 
32 & Entropic intensity of revised implication (IIER) & $ \sqrt{ \Big [ \big ( 1 - h_{1}(P(X\bar{Y}))^{2} \big )  \times \big ( 1 - h_{2}(P(X\bar{Y}))^{2}   \big )  \Big ]^\frac{1}{4} }$ $ \times  \sqrt{max(2 \times II - 1 ; 0)}$ \\ 
\hline 
33 & Likelihood discriminant index & $P \Big [ Poisson (n P(X) P(Y)) <  P(XY) \Big ]$ \\ 
\hline 
34 & Interest or Lift & $ \frac {p(XY)} {p(X)p(Y)} $ \\ 
\hline 
35 & Jaccard & $ \frac{p(XY)} {p(X\bar{Y}) + p(Y)}$ \\ 
\hline 
36 & J-Measure & $p(XY)~  log \Big (\frac{p(XY)}{p(X) p(Y)} \Big ) + p(X\bar{Y}) ~ log \Big (\frac{p(X\bar{Y})} {p(X) p(\bar{Y})} \Big ) $ \\ 
\hline 
37 & Klosgen &$ \sqrt{p(XY)}~ \Big( \frac {p(XY)} {p(X)} - p(Y) \Big )$ \\ 
\hline 
38 & Kulczynski or Agreement and disagreement index & $ \frac {p(XY)} {p(X\bar{Y}) + p(\bar{X}Y)}$ \\ 
\hline 
39 & Laplace & $\frac {n p(XY) +1} {n p(X) +2} $ \\ 
\hline 
40 & Leverage & $\frac {p(XY)} {p(X)} - p(X) p(Y) $ \\ 
\hline 
41 & $ M_{GK} $ & $  If~P(Y/X) \geq P(Y)~then~M_{GK}(X \rightarrow Y) = \frac {p(Y/X) - p(Y)} {1 - p(Y)} 
                     ~Else~M_{GK}(X \rightarrow Y) = \frac {p(Y/X) - p(Y)} {p(Y)} $ \\ 
\hline 
42 & Least contradiction or Surprise & $\frac {p(XY) - p(X\bar{Y})} {p(Y)} $ \\ 
\hline 
43 & Novelty & $ p(XY) - p(X)p(Y)$ \\ 
\hline 
44 & Pearl & $ p(X) | \frac {p(XY)} {p(X)} - p(Y) |  $ \\ 
\hline 
45 & Piatetsky-Shapiro & $n \times \Big( p(XY) - p(X) p(Y) \Big)  $ \\ 
\hline 
46 & Accuracy & $p(XY) + p(\bar{X}\bar{Y})$  \\ 
\hline 
47 & Prevalence & $ p(Y) $ \\ 
\hline 
48 & Yule's Q & $ \frac {p(XY) p(\bar{X}\bar{Y})  - p(X\bar{Y})  p(\bar{X}Y)}  
                             {p(XY) p(\bar{X}\bar{Y}) + p(X\bar{Y})  p(\bar{X}Y)} $ \\ 
\hline 
49 & Recall & $ \frac {p(XY)} {p(Y)}$ \\ 
\hline 
50 & Odds Ratio & $\frac {p(XY)p(\bar{X}\bar{Y})} {p(\bar{X}Y) p(X\bar{Y})} $ \\ 
\hline 
51 & Relative Risk & $ \frac {p(Y/X)} {p(Y/\bar{X})}$ \\ 
\hline 
52 & Sebag-Schoenauer & $ \frac {p(XY)} {p(X\bar{Y})}  $ \\ 
\hline 
53 & Specificity & $\frac {p(\bar{X}\bar{Y})} {p(\bar{X})} $ \\ 
\hline 
54 & Support or Russel and Rao index & $p(XY) $ \\ 
\hline 
55 & Yao and Liu's One Way Support & $\frac {p(XY)} {p(X)} log_2 \frac {p(XY)} {p(X) p(Y)} $ \\ 
\hline 
56 & Yao and Liu's Two Way Support & $p(XY) log_2 \frac {p(XY)} {p(X) p(Y)} $ \\ 
\hline 
57 & Examples and counter-examples rate & $\frac {p(XY) - p(X\bar{Y})} {p(XY)} $ \\ 
\hline 
58 & Test value VT100 & $ \phi^{-1} (P[Hypergeometric(100P(X)P(Y)) \leq P(XY)]) $ \\ 
\hline 
59 &  Yao and Liu's Two Way Support Variation & $p(XY)log_2 \frac {p(XY)} {p(X) p(Y)}  +
       p(X\bar{Y})log_2 \frac {p(X\bar{Y})} {p(X) p(\bar{Y})}  +
       p(\bar{X}Y)log_2 \frac {p(\bar{X}Y)} {p(\bar{X}) p(Y)}  +
       p(\bar{X}\bar{Y})log_2 \frac {p(\bar{X}\bar{Y})} {p(\bar{X}) p(\bar{Y})}  $ \\ 
\hline 
60 & Yule's Y & $ \frac {  \sqrt{p(XY) p(\bar{X}\bar{Y})}  -  \sqrt{p(X\bar{Y})  p(\bar{X}Y)}}
        {  \sqrt{p(XY) p(\bar{X}\bar{Y})}  +  \sqrt{p(X\bar{Y})  p(\bar{X}Y)}}$ \\ 
\hline 
61 & Zhang & $\frac {p(XY) - p(X) p(Y)}  {max \Big\{ p(XY) p(\bar{Y}) , ~ p(Y) p(X\bar{Y}) \Big\}  } $ \\ 
\hline 
%\end{tabular} 
\caption{Definition of 61 measures.} \label{tab:defmeasure}
\end{longtable}
\end{center}

%\section*{Author Biographies}
%\leavevmode
%
%\vbox{%
%\begin{wrapfigure}{l}{80pt}
%{\vspace*{15pt}\fbox{insert photo}\vspace*{100pt}}%
%\end{wrapfigure}
%\noindent\small 
%{\bf Li Shen} received a B.E. degree from Xi'an Jiao Tong University,
%Xi'an, China, in 1993 and an M.E. degree from Shanghai Jiao Tong
%University, Shanghai, China, in 1996.  }
%
%\vbox{%
%\begin{wrapfigure}{l}{80pt}
%{\vspace*{15pt}\fbox{insert photo}\vspace*{100pt}}%
%\end{wrapfigure}
%\noindent\small {\bf Hong Shen} is currently Associate Professor
%(Reader) in the School of Computing and Information Technology and
%Research Director of Parallel Computing Unit at Griffith
%University. He held visiting posts in several leading universities in
%USA, Europe and Asia. With main research interests in algorithms,
%parallel and distributed computing, interconnection networks, parallel
%databases and data mining, multimedia systems and networking, he has
%authored over 120 technical papers. }
%
%\vspace{15pt}
%\vbox{%
%\begin{wrapfigure}{l}{80pt}
%{\vspace*{15pt}\fbox{insert photo}\vspace*{100pt}}%
%\end{wrapfigure}
%\noindent\small 
%{\bf Ling Cheng} received a B.E. degree from Xi'an Jiao Tong
%University, Xi'an, China, in 1993 and an M.E. degree from Shanghai
%Jiao Tong University, Shanghai, China, in 1996. From 1996 to 1997, she
%worked in the IBM China Research Lab, Beijing, China.  }
%
%
%
%\correspond{Li Shen, Department of Computer Science, Dartmouth
%College, Hanover, NH 03755, USA. Email: li@cs.dartmouth.edu}
%\label{lastpage}
\end{document}